\documentclass[12pt]{article}
\usepackage{graphicx}
\begin{document}

\renewcommand{\thefootnote}{\fnsymbol{footnote}}
\renewcommand{\topfraction}{1}
\renewcommand{\textfraction}{0}

\title{\rightline{\rm\normalsize IIT-CAPP-15/2}
\Large\bf Introduction to Subatomic-
Particle Spectrometers\thanks{To appear in 
the Wiley {\it Encyclopedia of Electrical and Electronics Engineering}.}}
\author{Daniel M. Kaplan\\
{\sl Illinois Institute of Technology}\\
{\sl Chicago, IL 60616}\\
\qquad\\
Charles E. Lane\\
{\sl Drexel University}\\
{\sl Philadelphia, PA 19104}\\
\qquad\\
Kenneth S. Nelson\thanks{Now at Johns Hopkins University Applied Physics Laboratory, Laurel, MD 20723.}\\
{\sl University of Virginia}\\
{\sl Charlottesville, VA 22901}}
\bigskip
\date{}

\maketitle
\bigskip

\begin{abstract} 

An introductory review, suitable for
the beginning student of high-energy physics or professionals from 
other fields who may desire familiarity with subatomic-particle detection techniques. 
Subatomic-particle fundamentals and the basics of particle interactions with
matter are summarized, after which we review particle detectors. We conclude
with three examples that illustrate the variety of subatomic-particle
spectrometers and exemplify the combined use of several detection techniques to
characterize interaction events more-or-less completely. 
\end{abstract}

\newpage

\tableofcontents
\listoffigures
\listoftables



\newpage
\section{Introduction}

This article introduces the reader to the field of high-energy physics and 
the subatomic-particle detection techniques that it employs. 
We begin with an
overview of the field, then briefly introduce subatomic particles and their
detection before treating particle detectors in more detail.
We conclude with three examples that illustrate how a variety of detectors often work
together in  typical high-energy-physics experiments.

The experimental study of subatomic particles and their interactions has
revealed an unexpected layer of substructure underlying the atomic nucleus and
has shed light on the evolution of the universe in the earliest moments
following the Big Bang. This field of research is commonly referred to as
particle physics or (because of the highly energetic particle beams
often employed) high-energy physics.

Modern subatomic-particle experiments employ elaborate spectrometry systems,
often with state-of-the-art electronic instrumentation. 
While there is much variation among spectrometers, generally they measure 
the trajectories and energies of subatomic particles passing through them.
In a typical experiment, a beam of subatomic particles is brought into
collision with another particle beam or with a stationary target. 
Interactions between particles yield reaction products, some of which pass 
through the spectrometer. Measurements can include the 
momentum, angle, energy, mass, velocity, and decay distance
of reaction products. 

Particle detection techniques pioneered in high-energy physics have received
broad application outside of that field. Cases in point include nuclear
physics, astronomy, medical imaging, x-ray scattering, diffraction, and
spectroscopy, and the use of synchrotron radiation in biophysics, 
biochemistry, materials science, and the semiconductor industry
\cite{Mandelkern,Westbrook}. 
(There has of course been intellectual traffic in both 
directions, e.g., the pioneering use of semiconductor detectors in nuclear 
physics and of CCDs in astronomy~\cite{Janesick}.)

\section{{Overview of Subatomic Particles}}

Subatomic particles include the familiar electron, proton, and neutron
which are the components of the atom. In addition, dozens of less stable
particles have been discovered since the 1930s which can be produced in
reactions among electrons, protons, and neutrons and subsequently decay
in a variety of ways. Each particle is characterized by a unique set of
values for mass, electric charge, average lifetime, etc.
Subatomic
particles also possess a property called spin, a form of  of angular momentum that differs from the
classical concept in that it is quantized (in units of 
$\hbar/2$) and immutable. 
Table \ref{tab:units} defines the units of measure commonly used in 
high-energy physics for these quantities, which are the ones employed in 
this article.

\begin{table}[ht]
\setcounter{footnote}{0}
\begin{centering}
\caption[Units commonly used in high-energy physics]
{Units commonly used in high-energy physics}
\label{tab:units}
\vspace{5mm}
\begin{tabular}{lccc}
\hline\hline
Quantity & Unit & Value in MKS units$\footnotemark$ & Comment\\
\hline
\hline
charge & $e$ & $1.60\times10^{-19}$ C & \\
& electron-volt & & Kinetic energy of 
particle of charge $e$ \\
\raisebox{1.5ex}[0pt]{energy} & (eV) & 
\raisebox{1.5ex}[0pt]{$1.60\times10^{-19}$ J} & accelerated through 1 V. \\
mass & GeV/$c^2\footnotemark$ & 
$1.78\times10^{-27}$ kg & Mass and energy related by $E=mc^2$.\\
momentum & GeV/$c{\addtocounter{footnote}{-1}}\footnotemark$ 
& $5.34\times10^{-19}$ kg$\cdot$m/s & \\
&&&Reduced Planck constant;\\
\raisebox{1.5ex}[0pt]{spin} & \raisebox{1.5ex}[0pt]{$\hbar$} & 
\raisebox{1.5ex}[0pt]{$1.05\times10^{-34}$ J$\cdot$s} & 
spin quantum is $\hbar/2$.\\
\hline\hline

\end{tabular}
\end{centering}
\baselineskip=8 pt
\setcounter{footnote}{0}
{\footnotesize
$\footnotemark$
Values are quoted to three
significant digits, which is sufficient precision for most purposes.} \\
{\footnotesize
$\footnotemark$
1\,GeV = 10$^9$\,eV.}

\end{table}

\subsection{{Leptons, Hadrons, Gauge and Higgs Bosons}}

Since the 1960s a simple unifying scheme for the plethora of subatomic
particles has become generally accepted. Subatomic particles fall into four
categories: leptons, hadrons, and gauge and Higgs bosons (see Table \ref{tab:particles}). 
The hadrons are made of quarks (described below).
Leptons and quarks
each have one quantum of spin,
while gauge and Higgs bosons have an even number of spin quanta. 
Gauge bosons are responsible for the forces between 
particles. For example, the electromagnetic force arises from the exchange of
photons among charged particles, and the strong force from the exchange of 
gluons. Higgs bosons are believed to be a manifestation of the ``Higgs mechanism"\footnote{The more complete name, including  the authors who made the main contributions to the development of the idea, is the Englert--Brout--Higgs--Guralnik--Hagen--Kibble mechanism.} whereby fundamental particles receive nonzero masses owing to their interactions with a ``Higgs field'' permeating the universe.

\begin{table}[ht]
\setcounter{footnote}{0}
\caption[Properties of selected subatomic particles]{Properties of selected subatomic particles$\footnotemark\footnotemark$}
\label{tab:particles}
\vspace{5mm}
{\begin{centering}
\small
\begin{tabular}{l@{}c@{~}cccccc}
\hline\hline
&& &\bf Charge & \bf Mass$\footnotemark$\addtocounter{footnote}{-1} & \bf Mean life$\footnotemark$ & \bf Spin &\\
\raisebox{1.5ex}[0pt]{\bf Particle} && \raisebox{1.5ex}[0pt]{\bf Symbol} & 
$(e)$ & (GeV/$c^2$) & (s) & $(\hbar)$ &\\
\hline
\hline
{\bf Leptons:} &&&&&&&\\
\hline\hline
~electron && $e^-$ & $-1$& $5.11\times10^{-4}$ & stable & 1/2 & \\
~electron neutrino && $\nu_e$ & 0 & $<2\times10^{-9}$ & stable$\footnotemark$ & 1/2 & \\
~muon && $\mu^-$ & $-1$ & 0.106 & $2.20\times10^{-6}$ & 1/2 & \\
~muon neutrino && $\nu_\mu$ & 0 & $<2\times10^{-9}$ & stable{\addtocounter{footnote}{-1}}$\footnotemark$ & 1/2 & \\
~tau && $\tau^-$ & $-1$ & 1.78 & $2.90\times10^{-13}$ & 1/2 & \\
~tau neutrino && $\nu_\tau$ & 0 & $<2\times10^{-9}$ & stable{\addtocounter{footnote}{-1}}$\footnotemark$ & 1/2 & \\
\hline\hline
{\bf Hadrons:} &&&&&&&Quark content\\
\hline\hline
baryons:&&&&&&&\\
\hline
~proton && $p$ & +1 & 0.938 & stable & 1/2 & $uud$\\
~neutron && $n$ & 0 & 0.940 & 880 & 1/2 & $udd$ \\
~Lambda && $\Lambda$ & 0 & 1.12 & $2.63\times10^{-10}$ & 1/2 & $uds$ \\
&& $\Xi^-$ & $-1$ & 1.32 & $1.64\times10^{-10}$ & 1/2 & $dss$ \\
\raisebox{1.5ex}[0pt]{~cascade} & \raisebox{1.5ex}[0pt]{\Large\{}
& $\Xi^0$ & 0 & 1.31 & $2.90\times10^{-10}$ & 1/2 & $uss$ \\
~Omega && $\Omega$ & $-1$ & 1.67 & $8.2\times10^{-11}$ & 3/2 & $sss$ \\
\hline
mesons:&&&&&&&\\
\hline
& & $\pi^+$,$\pi^-$ & $+1,-1$ & 0.140 & $2.60\times10^{-8}$ & 0 &
$u\overline{d}$,$d\overline{u}$\\
\raisebox{1.5ex}[0pt]{~pion} & \raisebox{1.5ex}[0pt]{\Large\{} 
& $\pi^0$ & 0 & 0.135 & $8.5\times10^{-17}$ &
0 & $u\overline{u}$,$d\overline{d}$\\
& & $K^+$,$K^-$ & $+1,-1$ & 0.494 & $1.24\times10^{-8}$ & 0 & $u\overline{s}$ \\
\raisebox{1.5ex}[0pt]{~kaon} & \raisebox{1.5ex}[0pt]{\Large\{} 
& $K^0$,$\overline{K}{}^{0}$ & 0 & 0.498 &
$\footnotemark$ & 0 & $d\overline{s}$,$s\overline{d}$ \\
~J/psi && $J/\psi$ & 0 & 3.10 & $1.25\times10^{-19}$ & 1 & $c\overline{c}$\\
&& $B^+$,$B^-$ & $+1,-1$ & 5.28 & $1.64\times10^{-12}$ & 0 &
$u\overline{b}$,$b\overline{u}$\\
\raisebox{1.5ex}[0pt]{~$B$} & \raisebox{1.5ex}[0pt]{\Large\{} 
& $B^0$,$\overline{B}{}^{0}$ & 0 & 5.28 &
 $1.52\times10^{-12}$ & 0 & $d\overline{b}$,$b\overline{d}$ \\  
\hline\hline
{\bf Gauge bosons:} &&&&&&&$\!\!$Force mediated$\!\!$\\
\hline\hline
~photon && $\gamma$ & 0 & 0 & stable & 1 & $\!\!$electromagnetic$\!\!$ \\
& & $W^+$,$W^-$ & $+1,-1$ & 80.4 & $1.59\times10^{-25}$ & 1 & weak\\
\raisebox{1.5ex}[0pt]{~weak bosons} & \raisebox{1.5ex}[0pt]{\Large\{} 
& $Z^0$ & 0 & 91.2 &$1.32\times10^{-25}$ &
1 & weak \\
~gluons && $g$ & 0 & 0 & stable &1 & strong \\
~graviton$\footnotemark$
&& $G$ & 0 & 0 & stable & 2 & gravitational\\
\hline\hline
{\bf Higgs boson:}\\
\hline\hline
~Higgs && $H^0$ & 0 & 126 & $1.56\times10^{-22}\footnotemark$ & 0 & \\
\hline\hline
\end{tabular}
\end{centering}}
\break
\baselineskip=8 pt
{\footnotesize
\setcounter{footnote}{0}
{\footnotesize$\footnotemark$
Data presented here are for illustrative purposes; 
more complete and
detailed information is available in
the {\it Review of Particle Physics}~\cite{PDG}, published biennially and available on the World-Wide Web at http://pdg.lbl.gov.}\\
$\footnotemark$
Antiparticle symbols are indicated by an overline.\\
$\footnotemark$
Where available, values are quoted to three
significant digits, which is sufficient precision for most purposes.\\
$\footnotemark$
However, the neutrino species mix among themselves in flight (see text).\\
$\footnotemark$
Due to mixing of neutral kaons with their antiparticles, these 
particles do not have definite lifetimes. Symmetric and antisymmetric 
linear combinations of $K^0$ and $\overline{K}{}^0$, known as $K_S$ and $K_L$,
have lifetimes of $8.95\times10^{-11}$ s and $5.29\times10^{-8}$ s, 
respectively.\\
$\footnotemark$
The existence of this particle has been postulated but is not yet
definitively established.\\
$\footnotemark$
Predicted; measurements are not yet precise enough to determine the Higgs lifetime.}
\vspace{-0.2in}
\end{table}

Leptons and hadrons can be distinguished experimentally by their modes of
interaction. Hadrons are subject to the strong force\footnote{The strong and weak forces were formerly known as the strong and weak {\em nuclear} forces, but that usage has become less common as the realization dawned that they are more fundamental than the nucleus.} (which  binds the
nucleus together), while leptons are not. Only six types of
lepton are known: the (negatively charged) electron, muon, and tau and their
neutral partners, the electron neutrino, muon neutrino, and tau neutrino. 
While neutrinos interact with matter in these three definite {\em flavors}, 
they are observed to mix in flight, such that the flavor detected may be 
different from that which was originally emitted---a phenomenon commonly 
referred to as {\em neutrino oscillation}. Such behavior implies that the 
neutrinos are not massless (although the masses are too small to have been 
measured in experiments to date), and that the neutrino states of definite mass
(which are the ones that propagate in free space) are different from those of definite flavor.

The three charged leptons all have charge $-1e$. 
For each type of
lepton there exists a corresponding antiparticle. 
Lepton and antilepton have equal mass, spin, and lifetime, and 
electric charges equal in magnitude but (for
charged leptons) opposite in sign.

\subsection{{Neutrinos}}
Neutrinos are neutral leptons, paired with the charged
leptons (electron, muon, tau) or their antiparticles.
Since  neutrinos  interact with matter only via the weak force,\addtocounter{footnote}{-1}\footnotemark\ they are exceptionally difficult
to detect.  Experiments using neutrino detectors typically
employ intense sources of neutrinos or antineutrinos
(such as nuclear reactors, or neutrino beams from particle accelerators)
and massive detectors (1 to $>$\,1000 ton) in order to have
a sufficient  probability of neutrino interaction.

There are two known  mechanisms by which neutrinos
 interact with matter: 
\begin{enumerate}
\item Charged-current interaction: in which a neutrino  is converted
to its corresponding charged lepton, with a target particle
 contributing the electric charge; for example, an anti-electron-neutrino
interacting with a proton, producing a positron and neutron (this
 interaction is known as {\em inverse $\beta$ decay}).

\item Neutral-current interaction: in which a neutrino scatters from
another subatomic particle, transferring energy and momentum, but 
without conversion to a charged lepton.  
\end{enumerate}
Charged-current neutrino interactions are typically $\sim$\,100 times more probable
(depending on the target, energy, etc.),
so most neutrino detectors are designed for charged-current
 interactions,  inverse $\beta$ decay being 
especially common at lower energies.  The probability of 
a neutrino interacting with matter increases with energy, above
whatever threshold energy is required for the reaction to occur.

\subsection{{Quarks}}

The hadrons are composed of quarks, of which (like the leptons) 
only six types are known. These are
designated up and down, charm and strange, and top and bottom (see Table
\ref{tab:quarks}). (For historical reasons, the top and bottom quarks
are also designated by the alternative names truth and beauty; somewhat
illogically, top and beauty are the names more commonly used.) 
Like the leptons, the quarks come in pairs with the
members of a pair differing in electric charge by one unit. The up, charm,
and top quarks have charge +2/3$e$. The down, strange, and bottom quarks have
charge $-1/3e$. For each type of quark there exists a corresponding antiquark
with opposite electric charge.

\begin{table}[t]
\centering
\caption[The three generations of quarks and antiquarks]
{The three generations of quarks and antiquarks\setcounter{footnote}{0}\footnotemark}
\label{tab:quarks}
\vspace{5mm}
\begin{tabular}{lccccccc}
\hline\hline
Charge & Spin & \multicolumn{6}{c}{Generation:}\\
\multicolumn{1}{c}{($e$)} & ($\hbar$) & \multicolumn{2}{c}{1} & \multicolumn{2}{c}{2} & \multicolumn{2}{c}{3}\\
\hline
\hline
& &\multicolumn{6}{c}{Quarks:}\\
\hline
$+2/3$ & 1/2 && $u$ && $c$ && $t$\\
$-1/3$ & 1/2 & $d$ && $s$ && $b$ &\\
\hline
\hline
& &\multicolumn{6}{c}{Antiquarks:}\\
\hline
$+1/3$ & 1/2 & $\overline{d}$ && $\overline{s}$ && $\overline{b}$&\\
 $-2/3$ & 1/2 && $\overline{u}$ && $\overline{c}$ && $\overline{t}$\\
\hline
\hline
\multicolumn{2}{l}{Approx.\ mass (GeV/$c^2$):} & 0.005 & 0.002 & 0.095 & 1.3 & 4.2 & 170
\\
\hline\hline
\end{tabular}
\break
{\footnotesize
\addtocounter{footnote}{-1}
$\footnotemark$
After the ``Review of Particle Physics"~\cite{PDG}.\addtocounter{footnote}{-1}}
\end{table}

Quarks are bound together into hadrons by the strong force. This is primarily
observed to occur in two ways: a quark can bind to an antiquark to form a
meson or antimeson, and three quarks or antiquarks can bind together to
form a baryon or antibaryon. Bare quarks have never been observed and are
 presumed to be forbidden by the laws governing the strong force. 
The possible existence of hadrons made up entirely of gluons, four-
or five-quark 
states, or hadron--hadron ``molecules" is a subject of current experimental 
investigation but has not been definitively established.\footnote{Of course, the deuteron may be considered  a proton--neutron molecule.} 
 Some 
recently established states are apparently composed of two quarks
and two antiquarks, or possibly are ``molecular" bound states of two mesons
in close proximity. Similarly, apparent ``pentaquark" states might be five-quark states or baryon--meson molecules. These discoveries
may perhaps be the harbinger of a richer understanding of the bound states of quantum
chromodynamics (QCD, the theory of the strong force).

\section{{Overview of Particle Detection}}

Subatomic particles can be detected via their interactions with bulk matter. 
Most particles can interact via more than one of the four forces (in order of
decreasing interaction strength) strong, 
electromagnetic, weak, and gravitational.
It is typically the stronger forces that give the most dramatic and easily 
detectable signals. 
Since subatomic particles have such small masses, 
the gravitational force is entirely useless for their detection.
All charged particles can be detected via the electromagnetic 
force, since they ionize nearby atoms as they pass through matter.
Without exceedingly massive detectors, neutrinos (which as neutral leptons ``feel" only the weak and
gravitational forces) are exceedingly difficult to detect directly, and 
their production is
often inferred via conservation of momentum and energy by observing that
some of the momentum and energy present before a reaction are missing
in the final state.

\subsection{{Position Measurement: Hodoscopes and Telescopes}}

Detectors that measure particle position can be
arranged as {\em hodoscopes} or {\em telescopes}. Hodoscopes are arrays of adjacent
detectors typically used to measure the position of a particle along a
direction perpendicular to the particle's path. Telescopes are arrays of
detectors arranged sequentially along the particle's path so as to track the
motion of the particle. 

Commonly used position-sensitive detectors include scintillation counters,
solid-state detectors, proportional tubes, and multiwire proportional and drift
chambers. These produce electrical signals which can be digitized and processed
in real time or recorded for further analysis using 
computers. Specialized detectors less commonly used nowadays include the cloud
chamber and bubble chamber, in which measurements are made continually as the
particle traverses an extended gaseous or liquid medium, the spark chamber, and
stacks of photographic emulsion. These detectors typically produce information
on photographic film which must be processed optically, requiring scanning and
measurement by trained personnel or using sophisticated automated systems.

\subsection{{Momentum and Energy Measurement}}

\subsubsection{{Magnetic Spectrometry}}

In a magnetic field, charged particles
follow helical trajectories. The radius of curvature is proportional to 
particle momentum and inversely proportional to particle charge 
and field strength. Given the radius $r$ in meters, momentum $p$ in
GeV/$c$, charge $q$ in units of the electron charge, and field strength $B$ in
tesla, we have
\begin{equation}r=0.3\frac{p}{qB\sin{\theta}}\,,\end{equation}
where $\theta$ is the angle between the field direction and the particle 
momentum vector.
From measurements of the curvature of the particle track within the
field, the momentum can thus be determined. Even if no measurements are made
within the field, the curvature within it (and hence the momentum) can be
inferred by measuring the particle's
trajectory before and after it traverses the field. 
The
magnetic field, typically in the range one to
a few tesla, is generally produced using an electromagnet, which may be air-core
or solid and have conventional (copper or aluminum) or superconducting coils.

\subsubsection{{Calorimeters}}

Calorimeters are detectors of thickness sufficient to absorb as large a
fraction as possible of the kinetic energy of an incident
particle. While for electrons and hadrons this fraction can approach 100\%,
there is usually some leakage of energy out the back of a calorimeter.
An electrical signal is produced proportional to the
deposited energy. Unlike tracking detectors, calorimeters can detect neutral
as well as charged particles. Calorimeters also play an important role in 
electron identification and are sometimes used for muon identification, as
described next.

\subsection{{Particle Identification}}

Of the charged subatomic particles, five are sufficiently stable to travel many
meters at the energies typical in high-energy physics (one GeV to several 
hundred
GeV), so that their trajectories can be easily measured in a magnetic
spectrometer. The problem of particle identification is thus that of
distinguishing among these five particles: electrons, muons, pions, kaons, and
protons. In experiments that identify particles, multiple
particle-identification techniques are typically used together in order to
enhance the efficiency of identification and lower the probability of
misidentification.

\subsubsection{{Calorimetric Electron (and Photon) Identification}} 

As discussed in more detail in Secs.\ \ref{sec:radlength} and \ref{sec:calorim},
in material of high atomic number ($Z$), high-energy electrons create
characteristic electromagnetic showers consisting of a cascade of photons,
electrons, and anti-electrons (positrons). Thus the pattern of energy
deposition in a
calorimeter, as well as the correlation of deposited energy with
magnetically measured  momentum, can be used to distinguish electrons from other
charged particles. In a calorimeter optimized for this purpose, $e/\pi$
rejection of $10^{-4}$ can be achieved (i.e., only $10^{-4}$ of
pions mistaken
for electrons)~\cite{Appel} 
while maintaining $75$\% efficiency for electrons
(i.e., only 25\% of electrons rejected as having ambiguous identification)
\cite{Gaines}.

Since high-energy photons also create electromagnetic showers in high-$Z$
materials, electromagnetic calorimetry can  be used to identify photons and
measure their energy. Photons are distinguishable from electrons since they do 
not give observable tracks in tracking telescopes.

\subsubsection{{Muon Identification}}

Muons (and also neutrinos) are distinguished from other charged particles by
their low
probability to interact with nuclei: muons can pass through many meters of 
iron while depositing only ionization energy.
A muon can thus be identified efficiently and with little background,
with typical $\mu/\pi$ rejection of order $10^{-2}$~\cite{Murphy}, by 
its failure to shower in a calorimeter. Often for muon identification, instead
of a full calorimeter, a crude structure is used consisting of thick shielding
layers of steel or concrete insterspersed with detectors (an example
of such a muon-identification system is shown below in Fig.\ \ref{fig:HyperCP}).

\subsubsection{{Time of Flight and Ionization}}

\label{sec:TOFI}
At momenta up to a few GeV/$c$, particle velocity can be measured well enough 
for particle identification using time-of-flight measurement over a distance of 
order meters~\cite{D'Agostini}. 
This is typically accomplished using thick (several cm)
scintillation counters (see Sec.\ \ref{sec:scint}) to determine flight time to
a fraction of a ns. This information is often augmented by repeated
measurements of ionization rate in proportional chambers since (as
described in Sec.\ \ref{sec:eloss}) the rate of ionization in a medium is
velocity dependent.

\subsubsection{{Cherenkov Detectors}}
\label{sec:Cherenkov}

If a particle's momentum is known from magnetic spectroscopy, measurement of
its velocity determines its mass. The velocity can be measured (or limits can
be placed upon it) using the Cherenkov effect, by which a charged particle moving
through a transparent medium at a speed greater than the speed of light in that
medium emits photons at a characteristic velocity-dependent angle. (This
process is mathematically analogous to the emission of sonic boom by a
supersonic object or the creation of a bow wave by a fast-moving boat.) The
speed of light in a medium is slower than the speed of light in vacuum by
the factor $1/n$, where $n$ is the medium's refractive index.

Threshold Cherenkov counters~\cite{Cherenkov}
determine limits on a particle's speed by
establishing that the particle does or does not emit Cherenkov photons in
media of various refractive indices. Several threshold Cherenkov counters
with appropriately chosen thresholds can be used together to distinguish pions,
kaons, and protons within a given momentum range. This technique is typically
useful from about 1\,GeV/$c$ up to several tens of GeV/$c$. Ring-imaging
Cherenkov counters~\cite{RICH} measure the particle's speed by determining the
photon emission angle directly, and can be used up to a few hundred GeV/$c$.

Note that Cherenkov detectors are hardly useful for muon identification, since
muons and pions are so similar in mass that their Cherenkov thresholds (and 
photon emission angles) are nearly indistinguishable in practice.

\subsubsection{{Transition-Radiation Detectors}}

Transition radiation consists of photons emitted when a charged particle
crosses an interface between media of differing refractive indices. Particles
with {\em highly relativistic} velocity (i.e., with kinetic energy greatly
exceeding their mass energy) produce detectable numbers of {\em soft} x-rays 
(energy of order a few keV) when
traversing stacks of thin metal or plastic foils typically including 
hundreds of interfaces.
These x-rays can be detected in proportional chambers and used for $e/\pi$ 
discrimination at momenta exceeding 1\,GeV/$c$ and hadron ($\pi/K/p$)
identification up to a few hundred GeV/$c$~\cite{Andronic}.
Using calorimetry and transition-radiation detection together, $e$/$\pi$
rejection of $10^{-5}$ has been achieved~\cite{KTeV-e-pi}.

\subsection{{Neutrino Detection}}

Given sufficient detector mass, the direct detection of high-energy ($>$\,1\,GeV) neutrinos is similar to 
that of other particles of those energies. 
The neutrino must interact with matter in order to be detectable, after
which the products of the interaction  carry the majority of 
the neutrino energy and momentum. In some cases, the type 
of neutrino can be inferred from the interaction, such 
as when high-energy muons are produced in a muon-neutrino
interaction. 
\subsubsection{Reactor Neutrinos}
The first experimental detection of neutrinos was performed
by Reines and Cowan~\cite{Reines1953} 
in 1953 using  antineutrinos from the Savannah River nuclear reactor. Their
detector consisted of an organic liquid scintillator (see Sec.~\ref{sec:organic}) loaded with
cadmium, viewed  by  photomultiplier tubes to detect the scintillation
light. The liquid scintillator  had a significant hydrogen
content, which provided target protons for inverse $\beta$-decay
interactions, with the cadmium having a large probability of capturing
the resulting neutron and emitting gamma rays.  The neutrino 
event signature was a {\em prompt} signal from the positron 
kinetic energy and annihilation gammas, followed some $\mu$s
later by the {\em delayed} neutron-capture gamma signal.

The same general design of reactor antineutrino detector has been
in use since the early experiments---large organic scintillation
detectors with a neutron-capture isotope---although the choice
of isotope used to capture the neutron varies.
Both the prompt and delayed signals are detected as light pulses
from the scintillator, with light intensity related to the 
energy deposited in the scintillator. In many cases, the distribution of relative light 
intensity and arrival time among the various phototubes can be used to locate the neutrino
interaction vertex within the scintillator volume to cm precision. 
\subsubsection{Detection of High Energy Neutrinos}
At higher energies,  detectors optimized for muon
neutrinos consist of {\em active}  layers, such as 
plastic scintillators or proportional tubes, interspersed with layers
of absorber such as iron or rock.  Only high-energy muons are 
able to penetrate multiple layers of absorber, suppressing most
potential backgrounds. High-energy muons are also
present in cosmic rays; however, when muons are observed coming 
from the direction of (and with timing appropriate to) an 
accelerator-produced muon-neutrino beam, it is a clear indication of a beam-associated 
neutrino interacting to produce a muon.  

New detectors optimized for the detection of electron neutrinos include the NO$\nu$A
liquid-scintillator detector~\cite{NOvA}
and a number of liquid-argon detectors of increasing size~\cite{ArgoNeuT,MicroBooNE,ICARUS}. These minimize or eliminate
the absorber layers so as to optimize the sensitivity to the path-length and energy-deposition-pattern differences between  electrons (the neutrino oscillation signal in a muon-neutrino beam) and muons.

Underground neutrino experiments also observe high-energy muons
and electrons traveling ``downward'' or ``upward," from the interaction of neutrinos 
in matter above,  below, or within the detector. 
The ``up-going'' high-energy neutrinos detected in this
manner come from the interaction of cosmic rays with the
Earth's atmosphere on the other side of the Earth.  In this way, water-Cherenkov
detectors (such as Super-K~\cite{Fukuda:2002uc}
and IceCube~\cite{Collaboration:2011ym}) have been used to make significant 
contributions to the study of neutrino oscillations. 
Water-Cherenkov detectors typically consist of a large volume of
water (or ice) instrumented with photomultiplier tubes. The Cherenkov light
detected by the photomultipliers gives the particle direction
and a measure of particle identification and energy.

\section{{Probabilistic Nature of Particle Reactions}}

Since subatomic-particle spectrometers deal with the smallest objects we know
of, they are exposed directly to the statistical aspects of quantum mechanics and
the ``microworld." It is a striking feature of the laws of quantum mechanics
that they do not predict the outcome of individual particle reactions, but only
particle behavior on the average. Nevertheless, most aspects of particle
detection can be understood using classical physics, and quantum uncertainty 
is rarely a dominant contribution to measurement error.

\subsection{{Example 1: Elastic Scattering}}

As a first example, if we consider a proton colliding elastically
with another proton, classical physics predicts exactly the scattering angle as
a function of the impact parameter (the distance between the centers of the
protons measured perpendicular to the
line of flight). However, quantum mechanically the proton is described not as a
hard sphere with a well defined radius, but rather as a wave packet, with the
square of the wave's amplitude at any point in space giving the probability for
the proton to be at that location. The impact parameter in any given collision
is thus an ill-defined quantity. 
It is not necessary to delve into the mathematical
complexities of quantum mechanics to realize that in this situation the
scattering angle for a given encounter is a random and unpredictable quantity.
What the laws of quantum mechanics in fact predict is the {\em probability} for
a proton to scatter through any given angle, in other words, the
scattering-angle {\em distribution}.
The random nature of quantum mechanics underlies Heisenberg's famous 
uncertainty relations, which give the fundamental limits to the accuracy
with which any quantity can be measured.

\subsection{{Example 2: Inelastic Scattering}}

Next we consider an {\em in}elastic collision between two protons, in which one
or both protons emerge in excited states that decay into multiple-particle
final states. This is a common type of interaction event of interest in
high-energy
physics, since from the properties and probability distributions of the final
state can be inferred various properties of the protons, their constituent
quarks and gluons, and the interactions among them. In any given encounter
between two protons, whether an inelastic collision will take place cannot in
principle be predicted, nor, if so, what particles will be produced, and with
what momentum and spin vectors. What quantum mechanics {\em does} predict (in
principle) is the {\em distributions} of these quantities over a large number
of collisions.\footnote{However, since we do not yet have a completely satisfactory
theory of the strong force, these distributions cannot as yet be predicted in
detail from ``first principles."}

\subsection{{Classical Uncertainty}}

One should note that uncertainty in the outcome or measurement of an event is
often not quantum-mechanical in origin. For example, even classically, the
angle of elastic scattering of a given proton incident on a target is in 
practice not
predictable, since it is not feasible to measure the position of the proton
with respect to the scattering nucleus with sufficient precision to know the
impact parameter. Thus even a classical analysis of the problem predicts only
the scattering-angle distribution.

\subsection{{Measurement Resolution}}

When measurement yields a distribution for some parameter rather than
a definite
value, we can characterize the quality of the measurement by the width of the
distribution, i.e., the measurement {\em resolution} or uncertainty. 
Common ways of characterizing
the width of a distribution are the root-mean-square (rms) deviation and the
full-width at half-maximum (fwhm). Of course, if looked at in fine enough
detail, any measurement yields a distribution, though the distribution may be
extremely narrow in some cases.

A broad distribution in the result of some measurement can reflect
quantum-mechanical uncertainty or simply lack of knowledge of the exact input
state. We consider in Sec.\ \ref{sec:spect}
examples of both classical and quantum
contributions to measurement resolution.

\subsection{{Randomness and Experimental Instrumentation}}

For the designer of particle spectrometry systems, a consequence of these
uncertainties is that randomness must be taken into account. For example, one
might be designing data-acquisition equipment for an experiment intended to
operate at an event rate of 100\,kHz. This means that on average one has 10
$\mu$s to acquire and process the information from each event, but the
actual number of events occurring in a given time interval will be random and
characterized by a Poisson probability distribution (see Eq.\
\ref{eq:Poisson} below). Thus if a large fraction of all events are to be
captured, data acquisition for each event must be accomplished in a time that
is short compared to 10 $\mu$s (in this example), 
to keep sufficiently small the probability that
a second event occurs while the first is being processed.

\section{{Detailed Discussion of Particle Detectors}}

All detectors of subatomic particles operate by virtue of the energy lost by
charged particles as they traverse matter. Charged particles lose energy in
matter by several mechanisms. These include ionization of nearby atoms,
bremsstrahlung (emission of photons in the electric field of an atomic
nucleus), Cherenkov and transition radiation, and strong nuclear interactions.
For all charged particles except electrons, ionization typically dominates over
other mechanisms. The key challenge to particle detection is amplification of
the small signals (typically tens to thousands of photons or
electrons) produced by these mechanisms.

\subsection{{Ionization Energy Loss}}

\label{sec:eloss}

\begin{figure}
\centerline{\includegraphics[width=.6\linewidth,trim=140 210 180 125mm,clip]{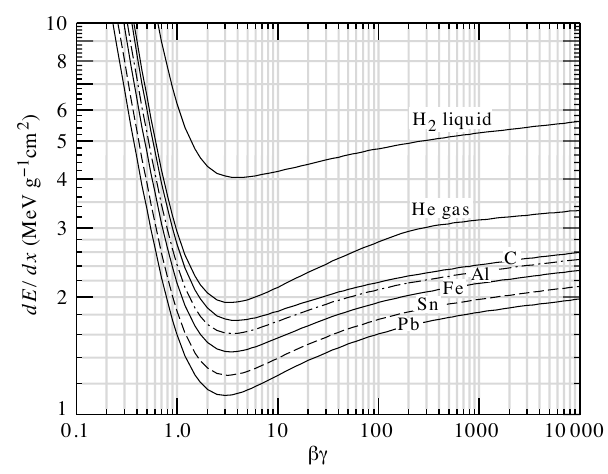}}
\caption[The
dependence of the ionization energy-loss rate on the relativistic speed
variable $\beta\gamma$ for particles of charge $e$ (except electrons) in
various materials]{The
dependence of the ionization energy-loss rate on the relativistic speed
variable $\beta\gamma$ for particles of charge $e$ (except electrons) in
various materials (after Fig. 22.2 of Ref.\ \protect\cite{PDG}). The ionization
rate first drops approximately as $(\beta\gamma)^{-5/3}$, then rises
logarithmically. Particles with $\beta\gamma>1$ (speed greater than
$c/\sqrt{2}$)
can loosely be 
considered ``minimum-ionizing."\label{fig:ionization}}
\end{figure}

The rate $dE/dx$ of ionization
energy loss by a charged particle passing through material depends 
primarily on the particle's speed, or more precisely on the quantity 
\begin{equation}\beta\gamma=\frac{v/c}{\sqrt{1-(v/c)^2}}\,,\end{equation} 
where $\beta\equiv v/c$ is the particle's speed expressed as a fraction of the
speed of light in vacuum, and $\gamma$ is the {\em time dilation factor}
$1/\sqrt{1-(v/c)^2}$.
(Note that
$\beta\gamma$ reduces to $v/c$ in the non-relativistic limit $v\ll c$, while in the ultra-relativistic limit  $v\gg c$, it approaches $p/mc$.)
The rate of ionization is given by the 
Bethe-Bloch equation; see Ref.~\cite{PDG} for details.
As shown in Fig.\ \ref{fig:ionization}, slow particles are heavily ionizing. 
The rate of ionization energy loss drops 
with increasing $\beta\gamma$ approximately as $(\beta\gamma)^{-5/3}$, to a
minimum at a value of $\beta\gamma$ that depends only slightly on the material.
For example, the ionization minimum is at $\beta\gamma=3.5$ for nitrogen, which decreases to
$\beta\gamma=3.0$ for high-$Z$ materials such as lead.

While the energy loss per unit thickness varies substantially among materials, if the
thickness is divided by density, thus being expressed as mass per unit area,
the strongest part of the dependence on
material is eliminated. For particles of charge $e$ (except electrons),
and for all materials except
hydrogen, the energy-loss rates at the ionization minimum 
range from 1 to 2\,MeV/(g/cm$^2$). For $\beta\gamma$ above
minimum-ionizing, the ionization energy-loss rate rises approximately
logarithmically.
At $\beta\gamma=10^4$, the loss rate is less than double relative to its
minimum at $\beta\gamma\approx 3$. 
In this ultra-relativistic regime radiative energy loss
(bremsstrahlung) becomes significant relative to ionization.

\subsection{{Radiation Length}}

\label{sec:radlength}

In materials of high atomic number, there is a high probability per unit length
for electromagnetic radiative processes to occur, i.e., for electrons to
radiate photons by bremsstrahlung and for photons (of sufficient energy) to convert into
electron--positron pairs in the electric field of a nucleus. This probability is
characterized by the {\em radiation length} $X_0$ of the material, defined as the
thickness of material in which a high-energy electron will lose all but a
fraction $1/e$ of its initial energy~\cite{Tsai}. Radiation length
also characterizes the degree to which charged particles scatter randomly, due 
to multiple encounters with the electric fields of nuclei, in 
passing through material. If precise measurement of particle trajectories is to
be achieved, this scattering effect must be minimized. Materials with short
radiation length (e.g., lead, $X_0=0.56$ cm, and tungsten, $X_0=0.35$ cm)
are thus desirable for use in electromagnetic sampling calorimeters (see
Sec.\ \ref{sec:calorim}) and in some shielding applications, but in general
should be avoided in other particle-detection contexts.

\subsection{{Scintillation Counters}}

\label{sec:scint}

Scintillators~\cite{Swank}
are materials in which some of the ionization energy lost by a
charged particle along
its trajectory is converted into light via fluorescence. The light may be
detected in a variety of ways, including (most commonly) photomultiplier tubes
and solid-state photodetectors. Both organic and inorganic scintillators are
in use.

\subsubsection{{Organic Scintillators}}
\label{sec:organic}
Organic scintillators typically consist of aromatic compounds, often dissolved in a
plastic (such as polystyrene, polyvinyltoluene, or polymethylmethacrylate) or an organic liquid (such as mineral oil or dodecane). Liquid scintillators have generally been used for ``large volume" (1--1000 ton) calorimetric detectors, where the cost of the material and light transmission over long distances are of primary concern. For smaller detectors, particularly with thin layers, plastic scintillators provide greater ease of use. A common configuration is a piece of plastic of few-millimeters' to few-centimeters' thickness, few- to several-centimeters' width, and length 
ranging from several centimeters to a few meters, glued at one end to a plastic
``light guide" which is in turn glued to, or butted against, the entrance window 
of a photomultiplier tube (PMT). 

A minimum-ionizing charged particle traversing  plastic deposits ionization
energy at a rate of about 2\,MeV/(g/cm$^2$). As the ionized plastic molecules
de-excite, they emit ultraviolet photons, most of which are quickly reabsorbed
by the plastic. To provide a detectable light signal, the plastic is
doped with a low concentration of dissolved aromatic {\em wavelength shifters} 
(fluors) such as $p$-terphenyl, 3-hydroxyflavone, or tetraphenylbutadiene.
These absorb in the ultraviolet and
re-emit at visible wavelengths, where the plastic is transparent. 
(Since there is inevitably some overlap between the wavelength-shifter
absorption and emission bands, too large a concentration of wavelength shifter would
result in excessive attenuation of the light signal as it travels towards the 
photodetector.)

In a counter of large length-to-width ratio, light collection is inherently
inefficient, since only a narrow range of emission angle is subtended by the
photodetector.
Furthermore, the light is attenuated by absorption along the length of the 
counter. Light is typically emitted at a rate of about 1
photon per 100\,eV of ionization energy, but often only a few percent of these
reach the photodetector,
where additional losses may be incurred due to reflection at the
interfaces.
The quantum efficiency of the photodetector further reduces the signal.
For a PMT, the quantum efficiency is the probability that an incident photon
causes the emission of an electron from the photocathode.
The typical PMT visible-light quantum
efficiency is about 20\%, but solid-state photodetectors can have quantum 
efficiencies approaching 100\%~\cite{Eberhard}.
Since photodetectors are subject to
single-electron shot noise, the typical signal-to-noise ratio in a
plastic scintillation counter of about 1 cm thickness is of order 10 to 100.

With fast fluors, the light signal develops quite rapidly (rise 
times of order ns).
Instantaneous counting rates of order 10\,MHz can be sustained. With
high-speed PMTs, thick scintillators can achieve sub-nanosecond timing
accuracy, ideal for time-of-flight particle identification. 
Average counting rates are limited by the current ratings of the
PMT and the {\em base} to which it connects and which provides its 
needed operating voltages. In high-counting-rate applications, transistorized
bases~\cite{Kerns} are crucial in order to avoid ``sagging" of the dynode voltages.

\subsubsection{{Scintillating Fibers}}

In recent decades advances in photodetectors and in the manufacture of plastic
optical fibers have made scintillating optical fibers a practical detector for
precision particle tracking at high rates~\cite{Ruchti}. 
Scintillating fibers work by
trapping scintillation light through total internal reflection. 
Since the fibers are typically less than 1 mm in diameter, detection of the 
scintillation signal is technically challenging:
the ionization signal is only of order $10^3$ photons and the light trapping 
efficiency of order 1\%.
To convert the
scintillation photons efficiently to visible light, wavelength shifters of
large {\em Stokes shift} (i.e., large separation between the absorption and
emission bands) are required so that they can be used in sufficiently high
concentration (of order 1\%) without excessive attenuation. If the light is
detected with
solid-state cryogenic VLPCs (visible-light photon counters)~\cite{VLPC},
advantage can be taken of their 80\% quantum efficiency, so that a
trapped-photon yield as low as several per minimum-ionizing particle suffices
for good detection efficiency (see Eq.\ \ref{eq:Poisson}). Fibers as narrow as
300 $\mu$m in diameter can then be used over lengths of meters~\cite{scifi}.
An advantage of the
large Stokes shift is operation in the green region of the visible spectrum (as
opposed to the blue of conventional scintillators), so that yellowing of the
plastic due to radiation damage in the course of a high-rate experiment has
only a slight impact on performance. 

\subsubsection{{Inorganic Scintillators}}

Inorganic scintillators include doped and undoped transparent crystals such as
thallium-doped sodium iodide, bismuth germanate, cesium iodide, barium fluoride, and lead
tungstate. They feature excellent energy resolution and are typically employed
in electromagnetic calorimetry (see Sec.\ \ref{sec:homog}).
Some notable recent applications~\cite{si-phot} have featured
silicon-photodiode readout, allowing installation in the cramped interior of
colliding-beam spectrometers as well as operation in high magnetic fields.

\subsection{{Proportional and Drift Chambers}}

Developed starting in the 1960s by Charpak {\it et al.}~\cite{Charpak68},
proportional and drift chambers have largely supplanted visualizing detectors,
such as bubble chambers, as the work-horse detectors of high-energy physics, due
to their higher rate capability and feasibility
of manufacture and operation in large sizes. 
They typically can provide sub-millimeter spatial resolution of charged-particle 
trajectories over volumes of several m$^3$~\cite{cands84}.  Installations of these
detectors are commonly realized as hodoscopic arrays of anode wires immersed in
a suitable gas mixture and arranged so as to detect the ionization energy
released when the gas is traversed by a charged particle.
The passage of the particle causes an electrical pulse on
the nearest wire, yielding,
with the simplest type of signal processing, discrete coordinate measurements
(i.e., the true position of the particle is approximated by the location of the
wire).
Continuous coordinate measurements can be achieved by more sophisiticated
signal processing that provides interpolation between anode wires. Although the
primary use of these detectors is  position measurement,
they also find use in particle identification, via the detection of 
transition radiation and  measurement of ionization rate ($dE/dx$).

\subsubsection{{Proportional-Tube Operating Principle}}

Many proportional-chamber arrangements have been devised. The simplest
conceptually is the proportional tube, in which a single thin anode wire is
operated at a positive potential (of order kV) with respect to a
surrounding conducting cathode surface. The tube is filled with a gas
suitable for detecting the particles of interest. For example, charged
particles are readily detected in a variety of mixtures of argon
with hydrocarbons, while detection of x-rays, e.g., from transition
radiation or crystal scattering, is more efficiently
accomplished using a higher-$Z$ gas (such as xenon) as the major component.
The exact choice of gas mixture also depends on such experimental 
requirements as rate capability, position resolution, detector operable
life-span~\cite{Kadyk}, etc.

\paragraph{Size of Primary Ionization Signal.} 
A minimum-ionizing charged particle traversing a proportional tube deposits only
a small
fraction of its energy in the gas, in a number of ionizing collisions
averaging about 0.5 to 5 per mm$\cdot$atm, 
depending on ion species and gas composition.
Due to the independent and random nature of the
collisions, they are characterized by Poisson statistics:
\begin{equation}
\label{eq:Poisson}
P(n)=\frac{\mu^ne^{-\mu}}{n!}\,,
\end{equation}
where $P(n)$ is the probability to produce $n$ ionizing collisions when the
mean number produced is $\mu$.
Furthermore, because of the wide range of energies imparted in 
these collisions, the yield of electron--ion
pairs is subject to large fluctuations~\cite{Cluster}. 
Consequently, amplification
electronics designed to detect the passage of minimum-ionizing particles
through the tube should be capable of handling the large dynamic range 
(typically exceeding 10) of these signals. 
In contrast, soft x-rays interact in the gas primarily via the
photoelectric effect, 
giving a narrower range of
signal size,
since the amount of ionization is more closely correlated with the x-ray energy.

\paragraph{Electron and Ion Drift.}
Under the influence of the electric field in the tube, the electrons and
positive ions produced by 
the initial interaction separate and drift toward the anode
and cathode, respectively. In the range of electric-field strength $E$ typically
found in proportional tubes, the average drift velocity $u^+$ of the positive
ions is proportional to $E$; it is often expressed in terms of the ion
mobility $\mu^+=u^+/E$. This proportionality results from competition between
two effects: acceleration of the ion by the electric field and randomization of
its direction by collisions with gas molecules.
A typical drift field $E=1$ kV/cm gives ion drift velocity 
in the range (0.5 to 2)\,$\times10^3$ cm/sec depending on
ion species and gas composition.

In weak electric fields, electrons are transported in a 
manner similar to that of positive ions.
However, in a sufficiently strong electric field, the
electron's wavelength $\lambda$, which decreases
in inverse proportion to its
momentum $p$ according to the deBroglie relationship $\lambda=h/p$, becomes
comparable to the size of molecular orbitals. (Here $h$ is Planck's constant.)
In this regime, the probability
per encounter for an electron to scatter off of a molecule has a strong
dependence on the electron momentum, displaying successive minima and maxima as
the momentum increases.
In many gas mixtures, the net effect is that the electron drift velocity
saturates, becoming approximately independent of electric field~\cite{Biagi89}.
This saturation typically occurs for fields in the neighborhood of 1 kV/cm, and
in argon-based mixtures results in a velocity of about 5 cm/$\mu$s.
The saturation of the electron drift velocity is an important advantage for
drift-chamber operation (Sec.\ \ref{sec:drift}) since it reduces the
sensitivity of the position measurement to operating conditions.

\paragraph{Development of the Avalanche Signal.}
As the electrons approach the anode wire, the electric field increases 
inversely with distance to the wire. Above an electric-field threshold 
whose value depends on the gas mixture, the electrons are accelerated between 
collisions to sufficient energy that they can ionize a
gas molecule on their next collision. Subsequently, the produced electrons
(along with the initial electron) are accelerated and produce further 
ionization. 
An avalanche multiplication of charge rapidly develops, 
with gain typically in the range $10^4$ to $10^6$ electron--ion pairs 
per initial electron.
Unlike the case of Geiger tubes and spark chambers, in proportional and drift
chambers the avalanche is normally not allowed to grow into a spark but
remains proportional in size to the amount of energy lost by the particle.
The avalanche develops essentially instantaneously (in a time interval 
$<1$ ns) within a few wire diameters of the anode. 

The time development of the anode current pulse is determined by the 
increasing separation of the electron--ion pairs generated in the avalanche. 
Using Green's reciprocity theorem~\cite{Griffiths} and assuming the 
anode is connected to a low-input-impedance amplifier, one can show that 
the increment $dq$ of induced charge on the anode due to the vector displacement
${\bf dr}$ of a charge $Q$ between the two electrodes is given by
\begin{equation}
\label{eq:Green}
dq = \frac{Q}{V_0} {\bf E}\cdot {\bf dr}\,,
\end{equation}
where $V_0$ is the anode--cathode potential difference and ${\bf E}$ the 
electric field.
(For cases with more than two electrodes, a more general
result can be obtained from the weighting field method~\cite{Ramo39}.)
At the instant of the avalanche, the electrons are collected on the
anode, giving rise to a sharp initial current pulse. However, since the
electrons are produced very near to the anode, this initial pulse
represents only a few percent of the total signal charge.
Thus it is the slow drift of the positive ions from anode to cathode
that provides most of the signal charge and determines the subsequent pulse
development. It should be noted that this imbalance between electron and
ion contributions to the signal is specific to the cylindrical 
proportional-tube geometry with thin anode wire. In a parallel-plate
geometry~\cite{Peisert}, the electrons contribute a much larger fraction of
the signal charge.

\subsubsection{{Multiwire Proportional Chambers}}

\begin{figure}
\centerline{\hspace{1. in}
\includegraphics{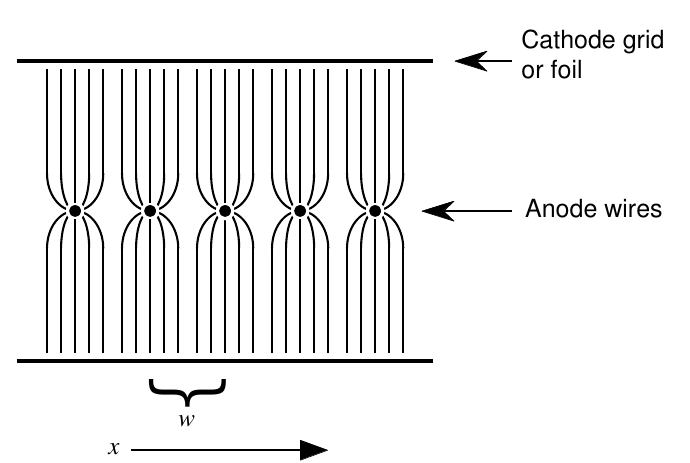}}
\caption[Sketch of the electric-field configuration in a
multiwire proportional chamber]{Sketch of the electric-field configuration in a
multiwire proportional chamber.
The anode wires are seen end-on. Close to the anode wire the field lines are 
radial as in a proportional tube, while close to the cathode planes the field 
lines are parallel as in a parallel-plate capacitor.
The presence of a signal on an anode wire determines the position of the 
particle along the $x$ axis in units of the anode-wire spacing $w$.
\label{fig:PWC-field}}
\end{figure}

In order to register the positions of many particles spread over some area, one
might employ a hodoscope made of individual proportional tubes.
In the simplest approach, the true position of the particle is then
approximated by the location of the wire that produces the pulse; i.e.,
measured positions are ``quantized" in units of the distance between adjacent 
anode wires.
Proportional-tube hodoscopes
have long been common in large-area, low-resolution applications
such as muon detection~\cite{Murphy,D0muon}, 
and have lately become popular in high-rate applications
as the straw-tube array~\cite{straw}. Another common
arrangement (which minimizes the detector material by eliminating the tube 
walls) is a multiwire
proportional chamber (MWPC), consisting of a planar array of anode 
wires sandwiched between two cathode foils or grids. 
Although in such a device the electric field near the cathodes 
approximates a parallel-plate configuration, close to an anode wire the field
shape resembles that in a proportional tube (see Fig.\ \ref{fig:PWC-field}).
The instrumentation of this detector usually involves
individual signal-detecting and coincidence/memory circuits for each anode 
wire. 

The strength of the MWPC is its ability to handle a high flux of
particles while providing good position resolution over a large detector area.
Rates above 10$^6$ particles per
cm$^2 \cdot$s have been achieved~\cite{Fischer85}
while maintaining greater than 95\% detection efficiency. 
The rate capability of an MWPC is a strong function of the anode-wire spacing 
and the anode--cathode gap width, both of which should be as small as possible
for high-rate operation.
The statistics of the primary ionization described in Eq.\ \ref{eq:Poisson}
implies a minimum anode--cathode gap of a few mm for 
efficient particle detection (i.e., for the
probability of no ionization to be negligibly small). In large detectors
the anode-wire spacing and anode--cathode gap are limited by electromechanical
instabilities~\cite{SFM}. Thus large MWPCs (anodes exceeding about 1 m in 
length) have typical anode spacing of a few mm, while for anode length under
10 cm, spacing down to 0.5 mm is feasible.
The anode-wire 
spacing $w$ determines the rms position resolution $\sigma$ of an MWPC  
according to
\begin{equation}
\sigma=\sqrt{\frac{1}{w}\int_{-\frac{w}{2}}^{\frac{w}{2}}x^2dx}=
\frac{w}{\sqrt{12}}\,. \label{eq:res}
\end{equation}
This resolution is not always achievable due to the interplay between the
pattern-recognition software and the occurrence of clusters of
two or more wires registering hits for a single incident particle.  
Such hit clusters can arise when two adjacent wires share the
ionization charge due to a track passing halfway between them, when several
adjacent wires share the charge due to an obliquely inclined track, or 
when an energetic {\em knock-on} electron (also known as a $\delta $ ray) 
is emitted at a large angle by the incident particle and traverses several 
adjacent wires.
When the best position resolution is required a proper
treatment of hit clusters is necessary, but not always possible,  leading
to inefficiencies in the track reconstruction.

\begin{figure}
\centerline{\includegraphics{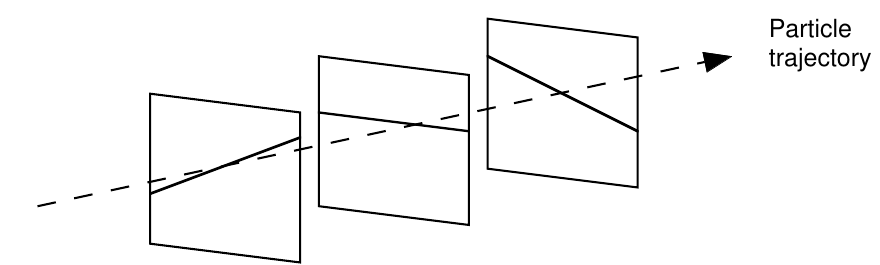}}
\caption[Schematic illustration showing how track positions in two dimensions
are determined from measurements in three successive MWPCs with anode wires at
three different angles]{Schematic illustration showing how track positions in two dimensions
are determined from measurements in three successive MWPCs with anode wires at
three different angles;
in each MWPC plane, only the wire producing a pulse is shown.
Two stereo views would be sufficient in principle, but
the third view helps to resolve ambiguities when more than one particle is
being measured simultaneously.
\label{fig:stereo}}
\end{figure}

Two shortcomings of the MWPC in measuring particle positions
are that the anode plane measures only one position coordinate (the one
perpendicular to the wire length),
and that the measured positions can assume only those values
corresponding to anode-wire positions. For minimum-ionizing particles,
a common way to provide information in the perpendicular coordinate direction 
(i.e., the distance along the wire) is
to employ several anode planes, each oriented at a different angle,
thus viewing the particle trajectory in {\em stereo} (Fig.\ \ref{fig:stereo}).
In a multiple-particle event, a software 
algorithm is then required to match up the signals corresponding to each 
particle in the various views.
In x-ray detection another method must be used, since x-rays
interact in the gas via the photoelectric effect
and are
thereby absorbed,
leaving a signal in only one plane.
A technique for two-dimensional position measurement using
only a single anode plane is charge division
\cite{Qdiv}, in which the ratio of charges flowing out the two ends of a
resistive anode wire specifies the position of the avalanche along the wire.
However, in practice this technique yields only modest resolution
(about 1\% of wire length).

\subsubsection{{Cathode Readout of Proportional Chambers}}

Another technique for two-dimensional measurement 
of particle position using only one plane of anode wires is cathode 
readout. Cathode readout also improves position resolution by allowing 
interpolation between anode wires.

\begin{figure}
\centerline{\hspace{.75in}
\includegraphics{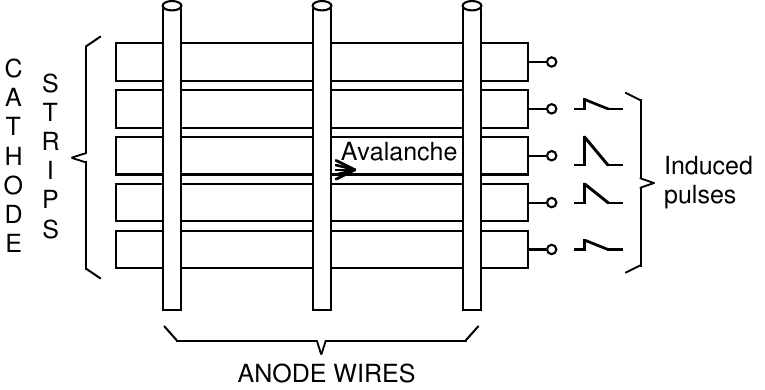}}
\caption[Schematic illustration of MWPC
cathode-readout principle]{Schematic illustration of MWPC
cathode-readout principle.
(Dimensions are not to scale but are exaggerated for clarity.)
The charges induced on the cathode strips, combined with the signals on the 
anode wires or on a second cathode plane (not shown) with strips parallel to
the wires, allow localization of the avalanche in two dimensions.
\label{fig:cathode}}
\end{figure}

When an avalanche occurs on an anode wire,
the pulse induced on the cathode planes carries information about the 
avalanche location.
If the cathode planes are segmented and the charge induced on each segment 
digitized, the avalanche location can be accurately determined from the 
center-of-gravity of the charge distribution.
The simplest arrangement 
is that of cathode-strip planes.
To obtain two-dimensional information, one cathode plane can have
strips oriented perpendicular to the wire direction (illustrated in Fig.\
\ref{fig:cathode}), and the other can have strips parallel to the wires.
Computation of the center of gravity
can be performed either ``off-line'' after the signals
have been digitized or ``on-line,'' for example 
via the transformation of signals from
the spatial domain to the time domain using delay lines and a specially patterned cathode
\cite{delayline}. The accuracy of such delay-line
position sensing is, in principle, independent of length. 

For cathode strips perpendicular to the wires, there is a small nonlinearity 
in position measurement due to the finite strip width~\cite{Gatti79}.  This
effect can be mitigated to some extent using specially shaped segmentation
patterns~\cite{Mathieson89}. For cathode strips parallel to the wires, the
measured position is  modulated by the discrete positions of the wires around
which the ions are created. However, at gains low enough that the avalanche
remains in  the proportional regime, the angular spread of the avalanche around
the  wire is limited~\cite{Harris78}, allowing some interpolation 
between the wires~\cite{Charpak78}. Typically the position resolution along the
wire is of order 100 $\mu$m and that perpendicular to the wire  approximately
five times worse.  

For cathode readout
measuring the coordinate along the wire, the accuracy of the center-of-gravity method is at least an order of magnitude 
better than that of resistive charge division (for a given signal charge).  
For the coordinate perpendicular to the wire the electron-drift timing
technique discussed in the next section is superior except when (as in x-ray 
detection) a timing reference is not available. 
The disadvantages of cathode readout are that
it requires a large number of well calibrated electronic channels and that
one is still faced with the problem of correlating the coordinate pairs in a
multiple-particle event. The latter problem is mitigated by the 
availability of a charge ``signature." Due to the large dynamic
range of primary ionization energy, cathode pulses from different
particles will tend to differ in total charge, while pulses from the same 
particle will be correlated in total charge in the two views.

\subsubsection{{Drift Chambers}}

\label{sec:drift}

One drawback of the MWPC is the large number of anode wires and
associated readout circuits needed to give fine position resolution 
over a large area. Drift chambers can substantially reduce the wire and 
electronics-channel counts. However, 
the generally wider anode spacing reduces rate capability, and the need for 
time measurement increases the electronic 
complexity of each instrumentation channel.

The idea is to record not only the position of the wire closest to the
particle trajectory, but also the distance of the particle from that wire, as 
measured by the time taken for the ionization electrons to drift along the 
electric-field lines in to the wire.
One thus records the time of occurrence of the avalanche relative to some
reference time tied to the passage of the particle. For
minimum-ionizing-particle detection, the reference time can be 
provided by a signal from a scintillation counter. 
Given the known drift velocity of the electrons,
the drift time determines the distance from the anode wire to an accuracy
(typically 100 to 200 $\mu$m) 
primarily limited by diffusion of the drifting electrons in the gas.  (In
special cases, such as high-pressure or specially tailored gas mixtures, the
diffusion contribution can be reduced so that the contribution from $\delta$
rays becomes the limiting factor, allowing sub-100 $\mu$m accuracy.)

\begin{figure}
\centerline{
\includegraphics{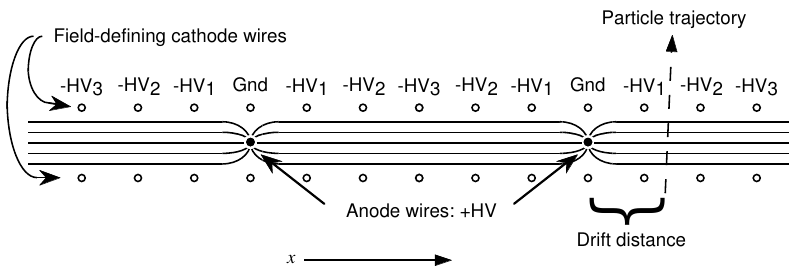}}
\caption[Sketch of the electric-field configuration in a drift chamber 
with ``graded" cathode potentials]{Sketch of the electric-field configuration in a drift chamber 
with ``graded" cathode potentials.
Wires are shown end-on.
The anode plane is sandwiched between two planes of cathode wires whose
high-voltages are stepped to produce an approximately constant drift field.
\label{fig:drift}}
\end{figure}

In a drift chamber the cathode planes are usually formed by wires at ``graded"
potentials, to provide a constant electric field for the drifting electrons
(Fig.\ \ref{fig:drift}). Use of a gas mixture having a saturated drift velocity
reduces the dependence of the
position measurement on field inhomogeneities and operating 
conditions. As in an MWPC, a single anode plane of a drift chamber measures
only the particle coordinate perpendicular to the wire direction. 
However, there is a two-fold (so-called ``left--right") drift-direction
ambiguity, which can be resolved by using several
planes of anode wires with positions staggered so as not to 
lie parallel to a possible particle trajectory. Although not often done, the
left--right ambiguity can also be resolved using the asymmetry of induced
charge on nearby electrodes,
assuming the avalanche
is sufficiently localized in azimuthal angle around the anode~\cite{Charpak73}.
The particle coordinate along the anode wire can be
obtained through similar means as in an MWPC: additional stereo-view
planes, charge division, or induced cathode signals.
Since for drift chambers, position
resolution is decoupled from anode-wire spacing, larger wire spacing than in
MWPCs is typically used, making them more straightforward to construct
and operate.

\subsubsection{{Time Projection Chambers}}

A time projection chamber~\cite{Nygren81} is a gas- (or, lately, liquid-)filled chamber in 
which ionization
electrons produced along the path of a charged particle drift over a
substantial distance (several cm to m) before avalanching and being 
detected in
an array of wire grids. With two-dimensional position measurement (e.g.,
anode and cathode readout), the entire particle trajectory through the chamber
is
recorded, with the third dimension ``projected" into drift time.
Such detectors are suitable when the average time between events is
sufficiently long compared to the drift time and when it is desirable to
identify particles via their $dE/dx$ ionization energy loss as discussed in 
Sec.\ \ref{sec:TOFI}.

\subsubsection{{Electronics for Proportional and Drift Chambers}}

Radeka~\cite{Radeka1,Radeka2} has analyzed the noise and resolution
considerations
in amplifiers for proportional chambers. 
Special operating conditions, such as high rates, bring additional concerns, 
some of which are discussed in~\cite{Boie}.

Since the majority of the signal charge stems from the slow motion of the
positive ions liberated in the avalanche, one can use Eq.\ \ref{eq:Green} to
show that the signal current has the form~\cite{Ricker}
\begin{equation}
\label{eq:ivst}
i(t) \propto \frac{1}{1+t/t_0}\,,
\end{equation}
assuming that the ions drift within a radial electric field directed outward
from the anode. The
characteristic time constant $t_0$ (of order 1 ns) depends on 
ion mobility, the electric-field strength at the anode, and the anode-wire 
diameter.

In high-rate applications it is desirable to cancel the slow ``$1/t$ tail"
in order to be ready for another incident particle as quickly as possible.
To do this, elaborate pulse-shaping circuits are sometimes employed using
multiple pole-zero filters~\cite{Boie}. With well designed amplifiers and pulse
shapers, typical double-pulse resolution of about 100 ns is feasible.
When designing such a circuit, one should
keep in mind that fluctuations (due to the arrival at the anode of individual
ions) with respect to the average pulse shape make perfect pulse-shaping
impossible. One must also consider that the amplifier
is to be connected to a wire operated at kV potential with
respect to a surface which may be only a few mm away. While sparks are highly
undesirable and may even break a wire, their occurrence cannot be ruled out.
Thus amplifiers should be provided with adequate input protection.

In typical MWPC or drift-chamber operation, the amplifier output 
(or the output of a separate pulse-shaping circuit if used) is conveyed to  a
``discriminator'' (a comparator driving a one-shot multivibrator) to produce a logic pulse
when an input exceeding threshold occurs.  The discriminator output may be used
to set a latch in coincidence with a reference ({\em gate}) signal, as is
typically done in MWPC installations.  The readout circuitry then provides a
list of wires having signals coincident with the gate. 
In drift-chamber operation one also needs to know the drift time of the
ionization electrons; in this case the time interval between the discriminator
output and the reference signal must be digitized. If cathode readout is
desired, the pulse heights on all cathode segments must also be digitized.

The amplifier, pulse-shaper, and discriminator may all be on separate 
multichannel circuit boards, combined on one or two circuit boards, or even 
all combined into a single hybrid or integrated circuit. Since for high-rate 
applications these circuits need to have both large gain and large bandwidth, 
and each is essentially connected
to an imperfectly shielded antenna (an anode wire or cathode strip), stabilizing large
installations against parasitic oscillation is usually challenging and
requires careful attention to grounding and circuit and detector layout.

Readout of the induced signals on cathodes usually requires longer
shaping time, and hence less bandwidth.  This is primarily due to
the longer time required for development of the signal, since 
the induced cathode charge increases as the ions drift
away from the screening influence of the nearby anode wire.
Nevertheless, since the accurate 
computation of the center of gravity requires a large dynamic range, to guard
against electromagnetic interference and cross-talk, the cautions just
mentioned concerning system layout apply here as well.

\subsection{{Solid-State Detectors}}

Silicon-strip detectors~\cite{Hall-rev} have come into increasing use for
tracking applications near the interaction vertex, where tracks are close
together and precise position measurements are needed. These detectors are
multiple-diode structures fabricated on single wafers of high-resistivity
silicon and operated under reverse bias. Center-to-center distance between
adjacent strips as small as 10 $\mu$m (25 to 50 $\mu$m is common) allows
position resolution at least an order of magnitude better than with
drift chambers.
The resolution achieved depends on readout mode: with single-bit-per-strip
digital readout (\'{a} l\`{a} MWPCs) the resolution is as given in Eq.\
\ref{eq:res}, while if analog pulse-height information is used, 
interpolation between strips is possible because of charge spreading over
adjacent strips; then rms resolution of a few $\mu$m can be achieved.

A charged particle traversing silicon creates electron--hole pairs
from ionization energy at the rate of one pair per 3.6\,eV. With typical 300 $\mu$m
detector thickness, the signal is about 25,000 electrons, an order of magnitude
smaller than in a proportional chamber. However, the smaller capacitance of a
silicon strip and its associated readout electronics
compared to that in an MWPC can allow improved noise performance.
This is especially true for silicon pixel detectors
\cite{Mouthuy,Kenney,Campbell}, in which an individual diode can have
dimensions of $30\times300$ $\mu$m$^2$ or less. Compared to strip detectors,
pixel detectors also offer ease of track reconstruction, since the
firing of a pixel determines a point in space along the particle trajectory
rather than a line segment.
To achieve efficient and rapid charge collection from the full thickness of the 
detector requires fully depleting the diodes, leading to typical operating 
voltage of about 100 V. The signal out of the $n$-type side
then develops in a few ns~\cite{Rudge}. 
With fast shaping time, extremely high particle rates (of order 
MHz per strip or pixel) can thus be 
handled, the limit to rate capability being radiation damage to the detectors 
and electronics over the long term.

Charged-coupled devices (CCDs) have also been employed for space-point tracking
close to the vertex~\cite{SLDCCD}. To achieve adequate signal-to-noise ratio
they must be operated with cryogenic cooling. CCDs have the virtue of good
position resolution ($<10$ $\mu$m rms) in both dimensions, at the expense of
long (of order 100 ms) readout time. They are thus not well suited to high-rate
experiments.

Other materials have been considered for strip and pixel particle-position 
measurement. At present much development effort is focused on the problem 
of radiation damage in vertex detectors~\cite{Hall}, since silicon detectors
commonly become unusable after a few megarad of irradiation. Due to their larger
band gaps, materials such as GaAs~\cite {GaAs} or diamond~\cite{diamond} 
are substantially more radiation-hard than silicon,
however they feature worse signal-to-noise ratio.

Reverse-biased silicon (and germanium) detectors and CCDs are also in
widespread use for x-ray and synchrotron-radiation detection~\cite{Westbrook},
nuclear physics,
etc.

Amplifiers and signal-processing circuitry for silicon-strip and pixel
detectors present challenges to the designer since the small feature size of
the detector implies very large channel counts (of order 10$^5$ strips or 
10$^8$ pixels) in an experiment. The cost per
channel is thus a key design criterion, and, since the circuits often need to
be packed into a small volume, so also are circuit size,
interconnections, and power dissipation. 
Various implementations have
lately been developed as semi-custom~\cite{ASIC} and full-custom integrated
circuits~\cite{SVX-II}. Pixel detectors necessarily require custom VLSI readout 
electronics, either integrated onto the detector chip itself~\cite{Kenney} or 
as a separate chip bump-bonded to the detector chip~\cite{Campbell}.

\subsection{{Calorimeters}}

\label{sec:calorim}

The term {\em calorimeter} as used here has been appropriated from chemical and thermal applications (with signals of microjoules or more) in order to describe detectors in which all or most of the energy (often picojoules or less) of a particle is absorbed so as to produce a measurable electrical signal.
Two common types of calorimeter are those optimized for the detection of 
electrons and photons (designated {\em electromagnetic}) and those optimized for 
strongly interacting particles (designated {\em hadronic}). Another important 
distinction is whether the output signal is proportional to all of the 
deposited energy or to only a portion of it; in the latter case the calorimeter
is of the {\em sampling} type.

\subsubsection{{Sampling Calorimeters}}

A common arrangement for a sampling calorimeter is a sandwich consisting of
layers of dense material interspersed with particle detectors such as
scintillation counters. Such a calorimeter can be electromagnetic or hadronic
depending on the dense material chosen. Often a combined
electromagnetic--hadronic
device is built, consisting of an initial electromagnetic section using lead
plates followed by a hadronic section using iron plates. Lead's short radiation
length (0.56 cm) combined with its long (17 cm) mean free path for hadronic
interaction means that electrons and hadrons can be well discriminated in such
a structure. Electrons interact in the lead producing an electromagnetic
shower as they radiate bremsstrahlung photons which produce electron--positron
pairs which in turn radiate photons, etc. Almost all of the electron's energy
is thus deposited in the electromagnetic section, which is typically about 20
radiation lengths thick. In a well designed calorimeter, the ionization energy
deposited
by the shower of electrons and positrons in the interspersed scintillator
active layers is proportional to the energy of the incident electron to
good approximation. Most hadrons pass through the electromagnetic section
leaving only ionization energy and proceed to interact strongly, producing a
hadronic shower, in the iron plates of the hadronic section.

Energy measurement in sampling calorimeters is limited in resolution due to
statistical fluctuations in the ratio of the energy deposited in the active
layers to that in the inactive ones. The percent resolution is inversely
proportional to the square-root of the deposited energy. Typical performance
for electromagnetic showers is relative rms energy uncertainty
$\sigma(E)/E=10\%/\sqrt{E}$ ($75\%/\sqrt{E}$ for hadronic), where $E$ is 
expressed in GeV. At the highest
energies, as this quantity tends to zero, other contributions (for example,
calibration uncertainties) dominate. It is difficult to measure energy
in sampling calorimeters to better than a few percent.

The poor energy resolution of hadronic sampling calorimeters arises from 
random fluctuations in the shower composition (e.g., in the relative numbers of
neutral vs.\ charged pions produced) and from energy loss mechanisms (such as
breakup of nuclei in the inactive layers) not yielding signal in the sampling
medium. The decay of the neutral pion into a pair of photons converts hadronic
energy into electromagnetic energy, which degrades the energy resolution due to
the differing response to electromagnetic and hadronic energy.
In {\em compensating} calorimeters, design parameters are tuned to minimize this 
response difference and thereby optimize hadronic energy resolution
\cite{Wigmans}.

Techniques for calibrating calorimeters include injecting light using lasers 
as well as studying the response to high-energy muons. 
Since muons do not shower, they deposit only 
minimum-ionizing pulse height in the active layers. The need to measure
with precision both muons and showers leads to stringent demands for 
analog-to-digital-converter linearity and dynamic range~\cite{KTeV}; 14 bits is
not uncommon.

\subsubsection{{Homogeneous Calorimeters}}

\label{sec:homog}

These include the inorganic scintillators discussed above as well as lead-glass
arrays and liquid-argon and liquid-xenon ionization chambers. Lead glass is
not a scintillator, but electrons and positrons from an electromagnetic shower
occurring within it emit visible Cherenkov light which can be detected using
PMTs. Large homogeneous calorimeters made with organic liquid scintillator 
are also used for neutrino detection, particularly for antineutrinos from nuclear reactors.
Since they are not subject to sampling fluctuations, homogeneous
electromagnetic calorimeters generally have better energy resolution than
sampling calorimeters, for example the 2.7\%/$E^{1/4}$ (fwhm) that was achieved
by the Crystal Ball collaboration using thallium-doped sodium iodide 
\cite{Xtal-Ball} and the 
5\%/$\sqrt{E}$ achieved by the OPAL collaboration using lead glass~\cite{OPAL}.
More recently the homogeneous scintillator PbWO$_4$ was developed for electromagnetic calorimetry in 
experiments at the LHC~\cite{PbWO4}.

\section{{Particle Spectrometers}}

\label{sec:spect}

Particle spectrometers feature great variety in their purposes and
layouts. Generically they may be divided into fixed-target spectrometers, in 
which a beam is aimed at a target that is stationary (or in the case of
a gas-jet or pellet target, moving slowly) in the laboratory, and colliding-beam
spectrometers, in which two particle beams moving in opposite directions are
brought into collision. We consider next three typical examples to illustrate the 
use of the detectors and techniques described above.

\subsection{{The Fermilab HyperCP Spectrometer}}

As a simple example of a fixed-target spectrometer we consider that of the
Fermilab HyperCP experiment~\cite{HyperNIM}. The goal of the
experiment was the precise comparison of decays of $\Xi^-$ baryons (quark
content $ssd$) with those of $\overline{\Xi}{}^+$ antibaryons ($\bar s \bar s
\bar d$), in order to search for a postulated 
subtle difference between matter and 
antimatter.  The difference in properties between matter and antimatter
had at the time been observed through the behavior of only one particle type (the $K^0$).\setcounter{footnote}{0}\footnote{~Since then it has  been observed in the behavior of  $B$ mesons as well.}
Nevertheless, it is believed  to be a general feature of the fundamental 
interactions among elementary  particles and, further, 
to be responsible for the dominance of matter over antimatter in the 
universe~\cite{CP}.

\begin{figure}
\vspace{-1in}
\centerline{
\includegraphics[width=\linewidth]{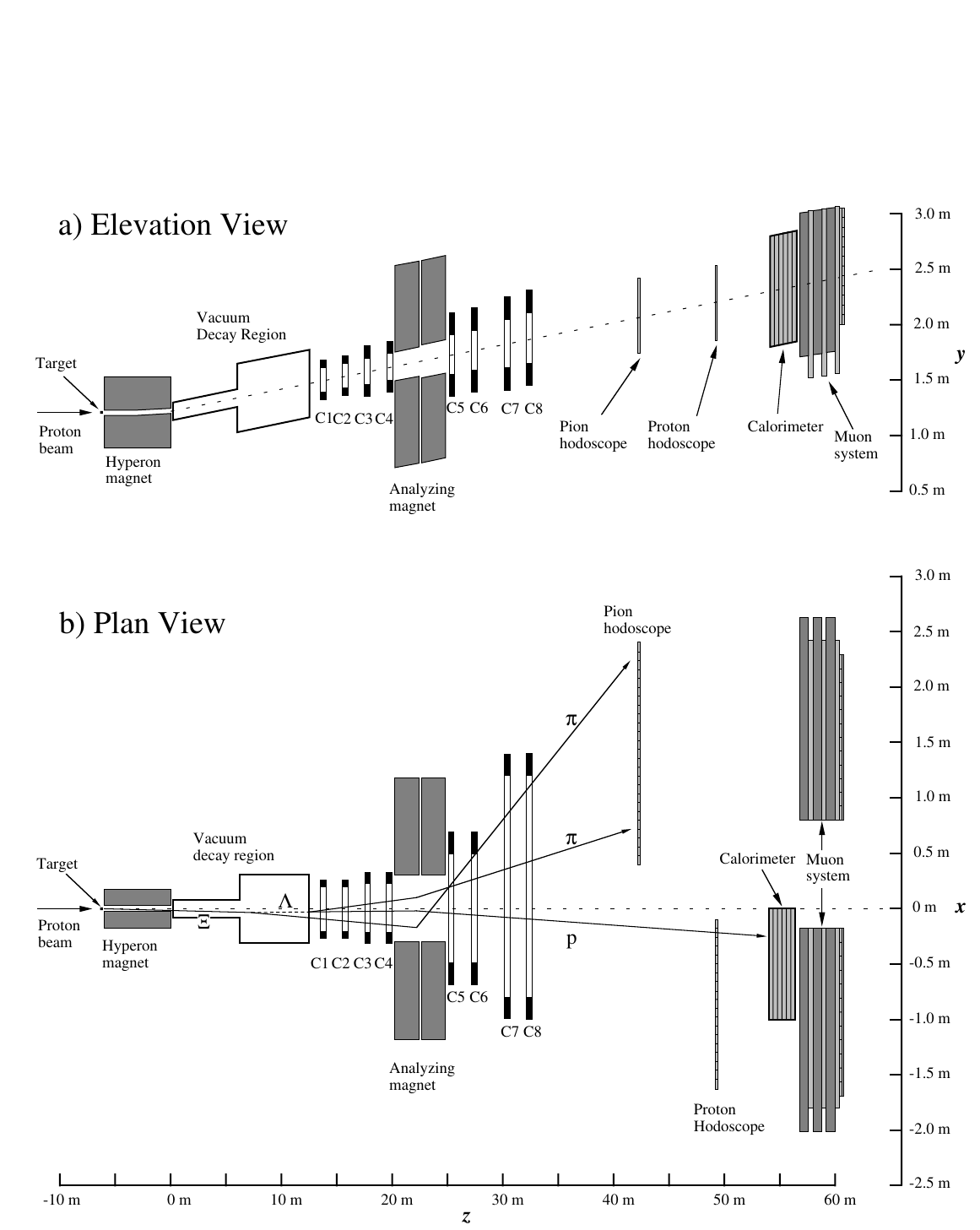}}
\caption[Elevation and plan views of the Fermilab HyperCP 
spectrometer]{(a) Elevation and (b) plan views of the Fermilab HyperCP 
spectrometer. (Note the different horizontal and vertical distance scales.)
Typical particle trajectories are shown for a cascade decay $\Xi\to\Lambda\pi$,
$\Lambda\to p\pi$. For simplicity, the curvature of the charged-particle tracks 
within the Analyzing magnet is approximated by a single sharp bend.
\label{fig:HyperCP}}
\end{figure}

Baryons containing strange quarks are known as {\em hyperons}. The $\Xi^-$ 
and $\overline{\Xi}{}^+$ hyperons were
produced by interactions of 800\,GeV {\em primary} protons from the Fermilab
Tevatron accelerator in a small metal target upstream of the ``Hyperon" magnet
(Fig.\ \ref{fig:HyperCP}). That magnet was filled with brass and tungsten
shielding,
into which a curved channel was machined such that charged particles of
momenta in the range 125 to 250\,GeV/$c$ traversed the channel and emerged out the
end to form the {\em secondary} beam;  neutral particles and charged particles outside that momentum range had trajectories that
curved either too little or too much and entered the shielding, where they showered
and were absorbed. 
The field directions in the Hyperon and Analyzing
magnets could be set to select either $\Xi^-$ or $\overline{\Xi}{}^+$ events.

\begin{figure}
\centerline{\hspace{-.125in}
\includegraphics[width=.5\linewidth]{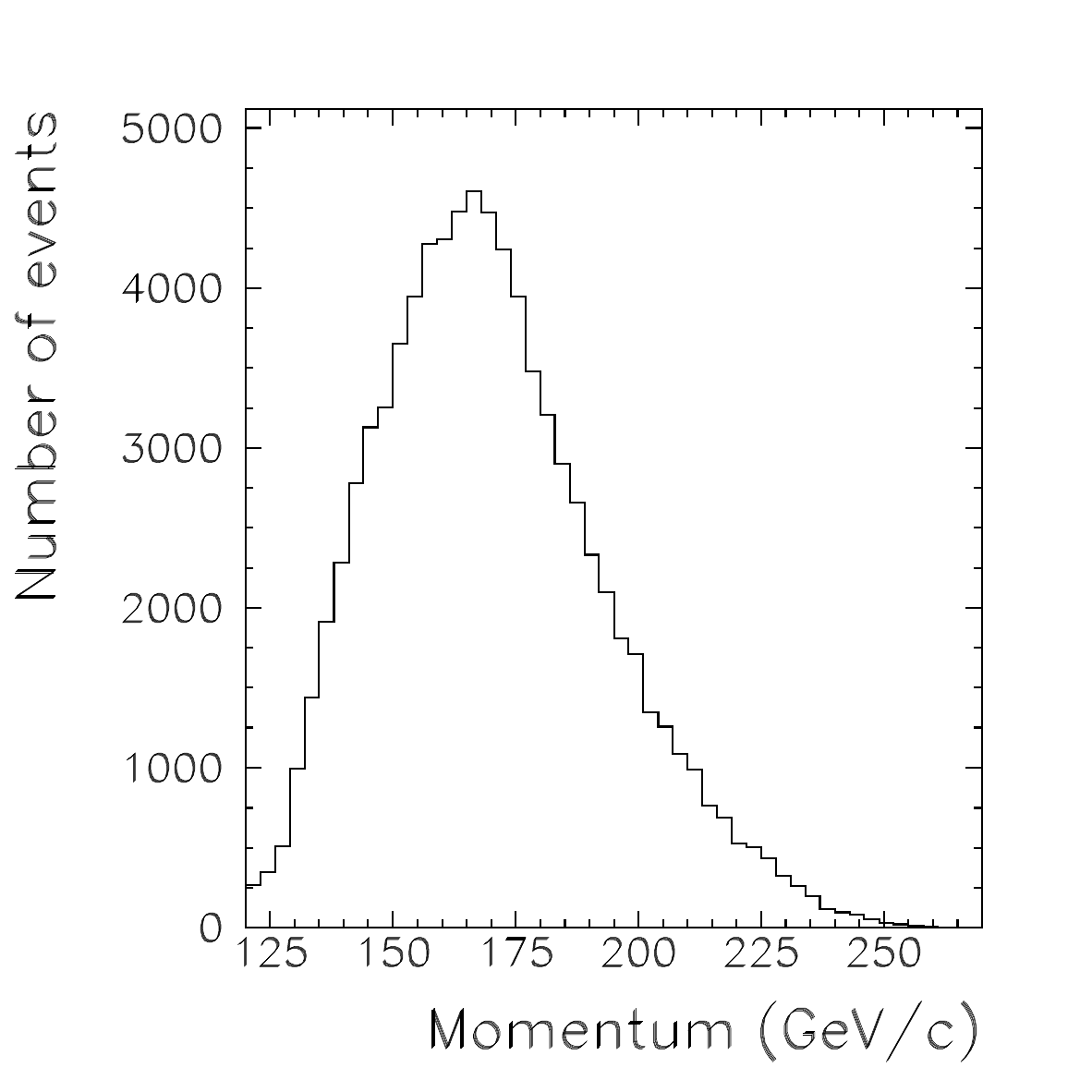}} 
\caption[Charged particles emerging from the HyperCP hyperon channel had
momenta distributed about a mean value of about 170
GeV/$c$]{Charged particles emerging from the HyperCP hyperon channel had
momenta distributed about a mean value of about 170
GeV/$c$.\label{fig:hycpenergy}} 
\end{figure}

Fig.\ \ref{fig:hycpenergy} shows the momentum distribution of charged particles
emerging from the channel in the positive-beam ($\overline{\Xi}{}^+$) setting. 
Note that this distribution arises classically, not quantum-mechanically:
To accept only a single value of momentum the channel would have needed to be 
infinitesimally narrow. Since its width was finite, it in fact accepted particles
over some range of track curvature and momentum.

The $\Xi^-$ or $\overline{\Xi}{}^+$
hyperon undergoes ``cascade" decay as each strange quark decays in
turn via the weak force. As indicated in Fig.\ \ref{fig:HyperCP}b, the $\Xi^-$
can decay into a
$\Lambda^0$ hyperon and a negative pion, and the $\Lambda^0$ can decay into a
proton and a negative pion. 
It is this decay chain that the HyperCP experiment studied.
The events of interest thus
contain a proton (or antiproton) of one charge and two pions of the opposite
charge. (For simplicity, $\Xi^-$ and $\overline{\Xi}{}^+$ are generically 
referred to simply as $\Xi$ in the following discussion, and $\Lambda^0$ and 
$\overline{\Lambda}{}^0$ as $\Lambda$.)

\subsubsection{{Triggering and Data Acquisition}}

The ``Pion'' and ``Proton'' hodoscopes (Fig.\ \ref{fig:HyperCP})
were arrays of vertical
scintillation counters used to trigger data acquisition from the
spectrometer. A trigger signal was created whenever counts were detected
simultaneously in both hodoscopes
and in the hadronic calorimeter. The state of all detector elements was then
digitized and recorded on magnetic tape for later computer analysis. The role
of the calorimeter was to suppress triggers that could occur when 
ambient muons or other low-energy particles counted in the hodoscopes. 
Large numbers of such background particles were
produced by particle interactions in the shielding, but their contribution to
the trigger rate was effectively suppressed by the calorimeter trigger
requirement. The 100\,kHz rate of event triggers was dominated by those interactions of
beam particles in the material of the spectrometer that gave counts in both
hodoscopes.

The HyperCP data acquisition system~\cite{HyperDAQ} had the highest throughput
of any then in use in high-energy physics. Digitization of event
information typically completed in $<3$\,$\mu$s, giving average {\em live time}
(the fraction of time that a system is available to process triggers) of
about 80\% at 100\,kHz trigger rate. To minimize the amount of information 
 recorded to describe each event, the spectrometer design was kept as
simple as possible, resulting in an average ``event size" of just 580 bytes.
Nevertheless the average data rate was about 15 MB/s, streamed to 40
magnetic tapes in parallel by 15 single-board computers housed in five VME
crates. (Since in fixed-target operation beam was extracted from the Tevatron
for only about 20 s each minute, the data acquisition rate from the digitizing
system was about three times the average rate to tape, with a 960 MB buffer
memory providing temporary data storage.) The 120 TB data sample thereby recorded  
was at the time  the largest of any high-energy-physics experiment. 

\subsubsection{{Coordinate Measurement}}

The trajectories of charged particles in the spectrometer were measured using a
telescope of multiwire-proportional-chamber modules (C1 to C8). Since 
the channeled secondary-beam rate exceeded that of $\Xi$ decays by a
factor $>10^{4}$, the rate capability of the these detectors was key to obtaining
the desired large sample (of order $10^9$ events) of hyperon and antihyperon
decays. To maximize rate capability, 1 mm anode-wire-spacing MWPCs were employed
for C1 and C2, with wire spacing ranging up to 2 mm for C7 and C8.
With a gas mixture of 50\% CF$_4$/50\% isobutane, 
C1 (which experienced
the highest rate per unit area) operated reliably at a rate exceeding 1\,MHz/cm$^2$.

To measure  particle positions in three dimensions, more than one
measurement view is required. Each of the eight chamber modules contained four
anode planes, two of which had vertical wires and two of which had wires at
angles of $\pm27^\circ$ with respect to the first two. 
This choice of stereo angle was found to optimize the measurement resolution
for hyperon mass and decay point.
Measurements were thus
provided in $x$ as well as in directions rotated by $\pm27^\circ$ with respect
to $x$, from which $y$ coordinates could be computed. $z$ coordinates were
given by the known locations of the MWPC planes. 

\subsubsection{{Event Reconstruction}}

Given the information from the MWPC telescope, three-dimensional reconstruction
of the momentum vector of each charged particle could be carried out on a
computer. Since momentum is conserved, the vector sum of the momenta of the
$\Xi$ decay products must equal the momentum vector ${\bf p}_\Xi$ of the $\Xi$
itself, and
likewise, since energy is conserved, the sum of the energies of the decay
products must equal the energy $E_\Xi$ of the $\Xi$. From the relativistic
relationship
among mass, energy, and momentum, the mass of the $\Xi$ could then be reconstructed as
\begin{equation}m_\Xi c^2=\sqrt{E_\Xi^2-p_\Xi^2c^2}\,.\end{equation}

This calculation requires knowledge of the energies of the $\Xi$ decay products.
To calculate the energy of a decay product from its momentum, its mass must be
known. While the masses could have been determined using Cherenkov counters, we 
will see below that in this instance it is sufficient simply to assume that the
two equal-charged particles are pions and the particle of opposite charge is
the proton or antiproton.
Of course, these assumed particle identities were not always correct,
nor were all observed combinations of a proton (or antiproton) and two 
negative (or positive) pions in fact decay products of a $\Xi^-$ (or
$\overline{\Xi}{}^+$). 

\begin{figure}
\centerline{\hspace{-.125in}
\includegraphics[width=.5\linewidth]{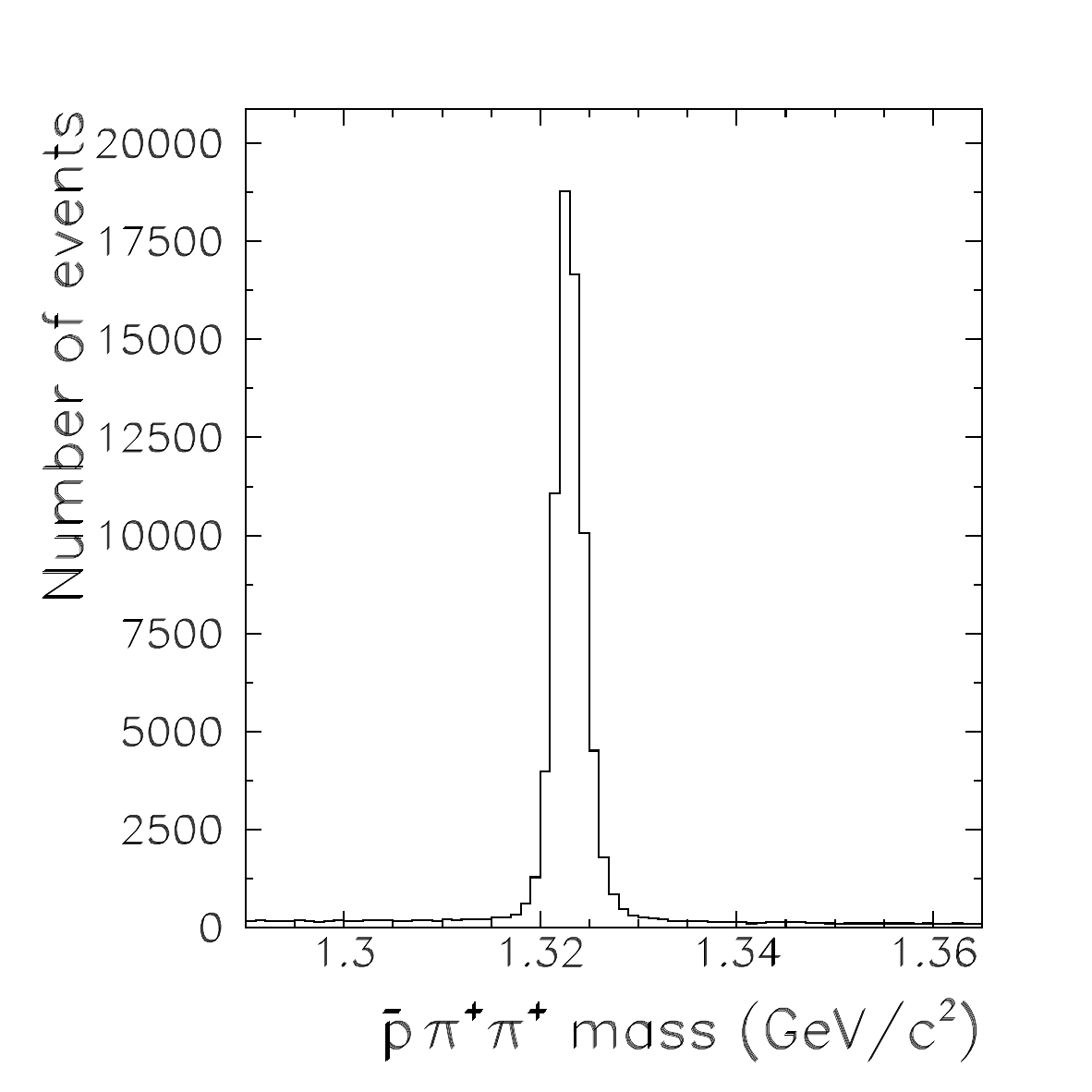}}
\caption[Distribution in effective mass of a sample of $\overline{p}\pi^+\pi^+$
combinations from the HyperCP experiment]{Distribution in effective mass of a sample of $\overline{p}\pi^+\pi^+$
combinations from the HyperCP experiment.\label{fig:ximass}}
\end{figure}

Fig.\ \ref{fig:ximass} shows the distribution in mass of a
sample of $\overline{p}\pi^+\pi^+$ combinations from the HyperCP experiment. A
clear peak at the mass of the $\overline{\Xi}{}^+$ is evident, superimposed on
a continuum of
background events in which the assumed particle identities were incorrect or the
$\overline{p}$, $\pi^+$, and $\pi^+$ not due to $\overline{\Xi}{}^+$ decay.
The width of the peak reflects 
uncertainties in the measurement of the particle momentum vectors.
These arise
from the wire spacings of the MWPCs and from  scattering of the
particles as they pass through the material of the detectors (again an example 
in which the 
measurement uncertainty is not quantum-mechanically dominated). By
requiring the reconstructed mass to fall within the peak, one could select
predominantly signal events and suppress background. The 
signal-to-background ratio could be improved by carrying out constrained fits to 
the cascade decay geometry and constraining the momentum vectors of the
$\Lambda$ decay products to be consistent with the known $\Lambda$ mass.
While the signal-to-background ratio could have been further improved by using 
Cherenkov counters for particle identification, the improvement would come at 
the expense of increased cost, complexity, and event size and was not needed 
for the purposes of the experiment.

\begin{figure}
\centerline{\hspace{-.25in}
\includegraphics[width=.5\linewidth]{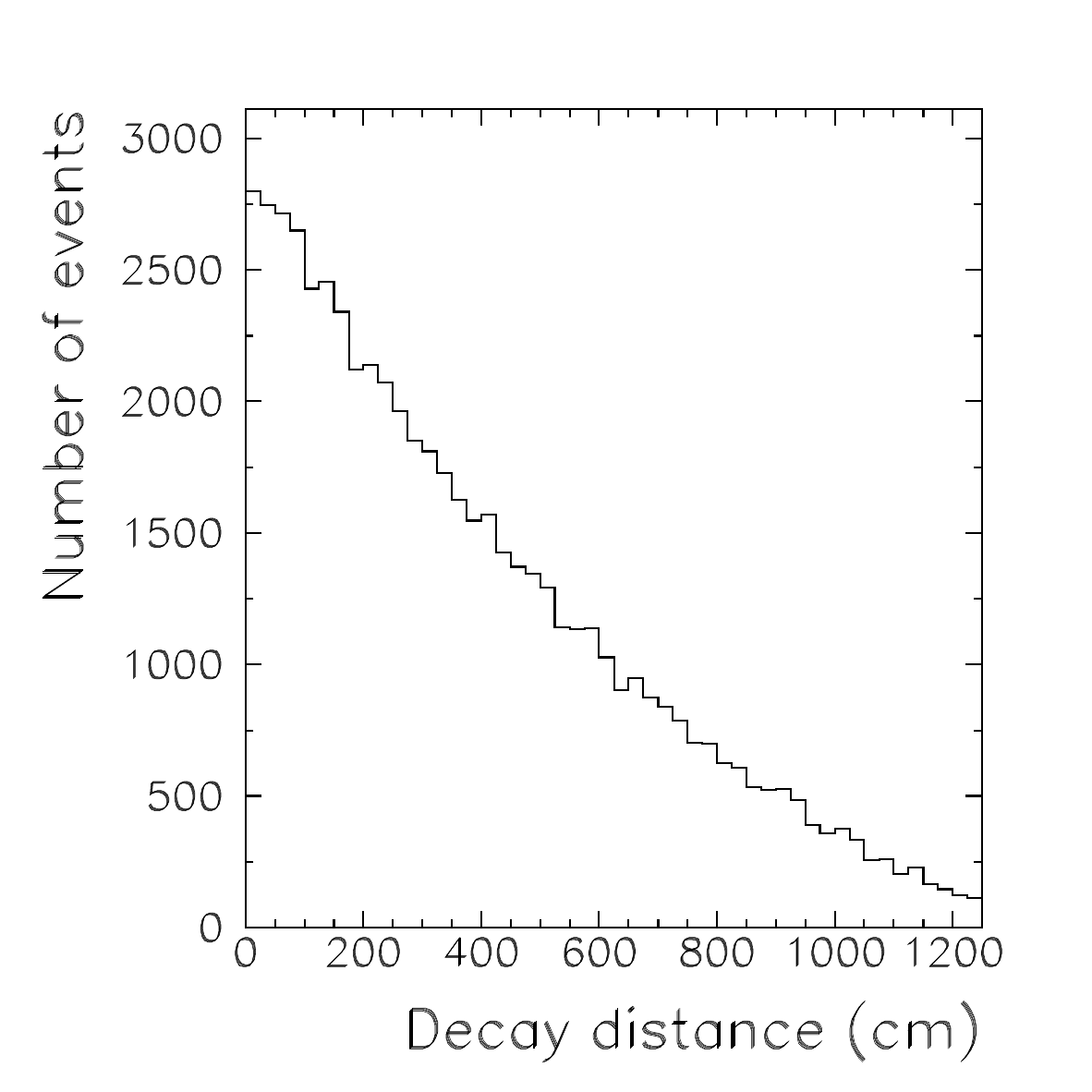}}
\caption[Distribution in decay distance of a sample of
$\overline{\Xi}{}^+\to\overline{\Lambda}\pi^+\to\overline{p}\pi^+\pi^+$
combinations from the HyperCP experiment]{Distribution in decay distance of a sample of
$\overline{\Xi}{}^+\to\overline{\Lambda}\pi^+\to\overline{p}\pi^+\pi^+$
combinations from the HyperCP experiment. To enhance the signal relative to the
background, the $\overline{p}\pi^+\pi^+$ mass is required to fall within $\pm$5
MeV/$c^2$ of the
known $\overline{\Xi}{}^+$ mass.\label{fig:xizvtx}}
\end{figure}

Also of interest in HyperCP was the distribution in decay point of the $\Xi$
hyperons. This is shown in Fig.\ \ref{fig:xizvtx}. The $\Xi$ decay point was
reconstructed for each event by first locating the point of closest approach of
the $\Lambda$ decay products.
This point approximates the position at which the $\Lambda$ decayed.
The $\Lambda$ trajectory was then extrapolated
upstream to its point of closest approach with the pion track from the $\Xi$
decay,  approximating the position at which the $\Xi$ decayed. 
An (approximately) exponential distribution is evident, as expected for
the
decay of an unstable particle. This reflects quantum-mechanical randomness: 
although the $\Xi$ has a definite {\em average} lifetime (as given in Table
\ref{tab:particles}),
the actual time interval from creation to decay of a given individual $\Xi$
cannot be predicted but varies randomly from event to event.
The deviations from exponential character arise
from three sources: (1) some background events are present in the sample, (2)
no correction was made here for the (momentum-dependent) relativistic time-dilation
factor $\gamma$, which was different for each event,
and (3) the detection probability was not entirely uniform for
$\Xi$ hyperons decaying at different points within the vacuum decay region.
These effects are all correctable in a more sophisticated analysis, but the
simple analysis presented here serves to illustrate the key points.

\subsection{{Collider Detector at Fermilab}}

To indicate the wide range of possible spectrometer configurations, we next
consider briefly the Collider Detector at Fermilab (CDF) spectrometer. This
is an example of a colliding-beam spectrometer notable for its use (along with
the D{\O} spectrometer) in the 1995 discovery of the top quark. 
A key difference between fixed-target and colliding-beam spectrometers is that 
in the former case the reaction products emerge within a narrow cone 
around the beam direction, whereas two beams colliding head-on produce reaction 
products that emerge in all directions. This leads to rather different 
spectrometer layouts in the two cases.
A typical design goal of colliding-beam spectrometers
is {\em hermeticity}, i.e., as few as possible of the particles produced in the
collisions should escape undetected. This of course contradicts the
requirements that the beams  enter in order to collide, the detectors be 
supported in place, and that the signals be
brought out; thus compromises are necessary. 

\begin{figure}[tb]
\centerline{\hspace{-.2in}
\includegraphics[width=\linewidth,trim=30 500 190 10mm,clip]{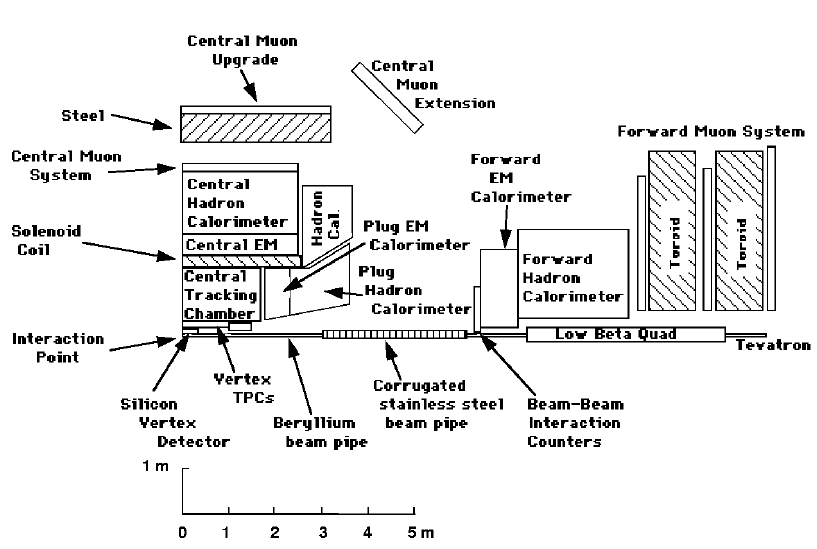}}
\caption[Schematic diagram of the Collider Detector at Fermilab
spectrometer]{Schematic diagram of the Collider Detector at Fermilab  
spectrometer. One-quarter is shown; the rest is implied by rotational symmetry 
about the beam pipe and mirror symmetry about the plane through the interaction
point perpendicular to the beam pipe.\label{fig:CDF_schematic}}
\end{figure}

\begin{figure}[tb]
\centerline{
\includegraphics{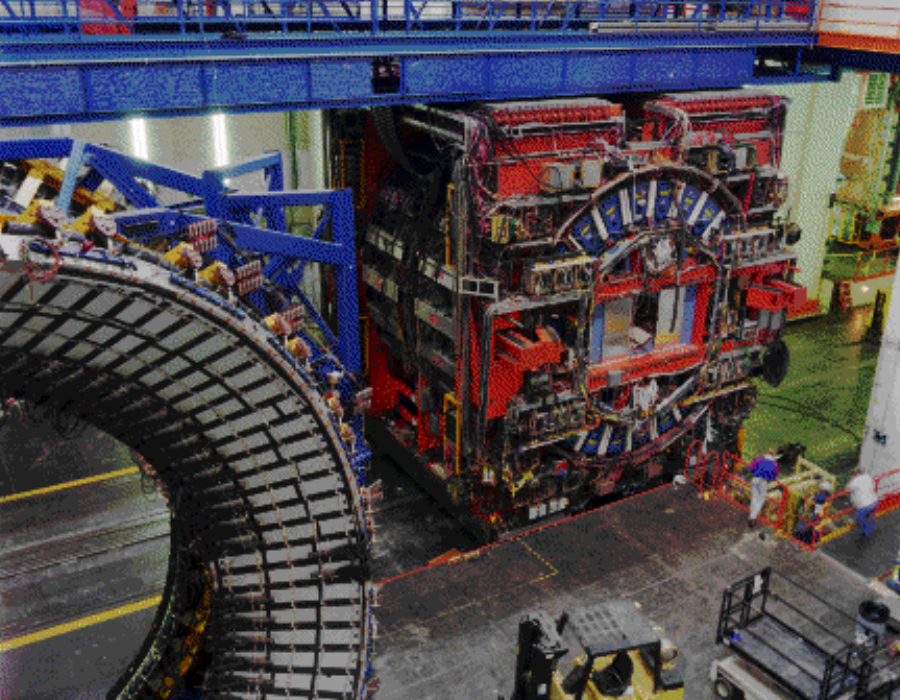}}
\caption[Photograph of the Fermilab Collider Detector Facility (CDF)
spectrometer in its assembly hall]{Photograph of the Fermilab Collider Detector Facility (CDF)
spectrometer in its assembly hall; the forward detectors have been retracted to
give access to the central portion.\label{fig:CDF_photo}}
\end{figure}

\begin{figure}[htb]
\centerline{
\includegraphics{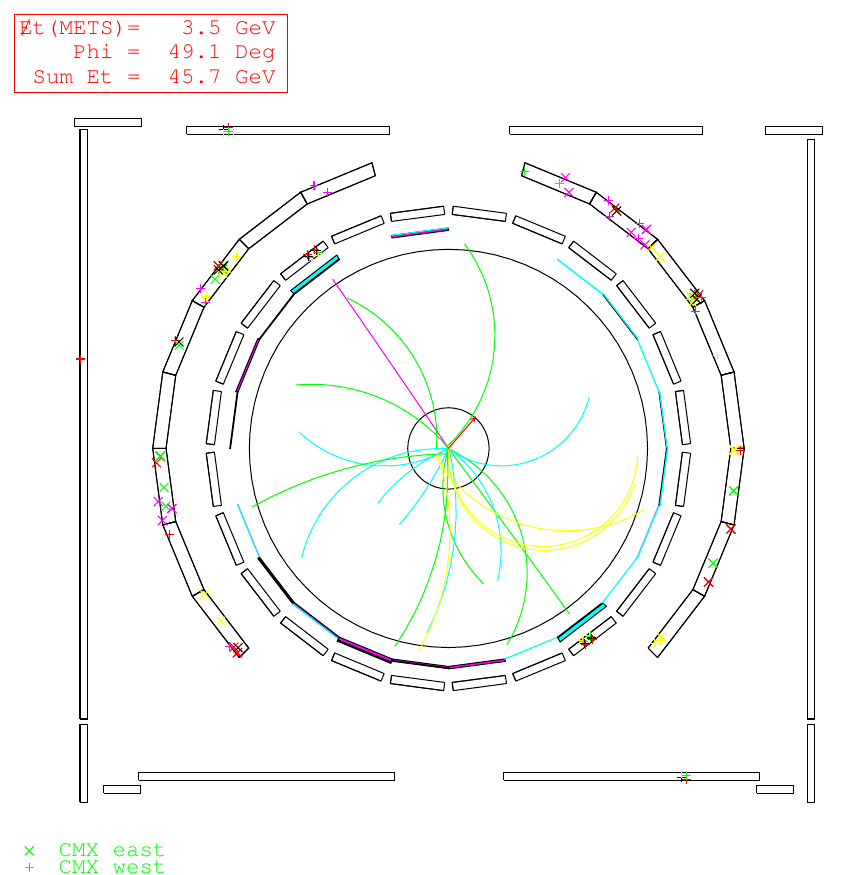}}
\caption[End-view display of a CDF event containing a $Z^0\to\mu^+\mu^-$ decay]{End-view display of a CDF event containing a $Z^0\to\mu^+\mu^-$ decay.
The muon tracks are the two line segments emerging back-to-back from the
interaction point at about 5 o'clock and 11 o'clock. They are identified as
muons by the ``X"'s that indicate signals in the inner and outer muon
detectors. Because of their high momentum
($p_{\mu}=m_{Z^0}c/2=45.6$\,GeV/$c$), the muon tracks show little curvature as 
compared to the tracks of the remaining (lower-momentum) charged particles in
the event. It is apparent that more tracks point down and to the left than up 
and to the right, suggesting that noninteracting electrically neutral particles
(neutrinos) may have been produced, or that some neutral particles were missed
due to cracks in the calorimeters. The ``missing momentum" vector due to the
undetected neutral
particles is indicated by the arrow.\label{fig:Zmumu}}
\end{figure}

Like the HyperCP detector, the CDF detector has
been described in detail in the literature~\cite{CDF-NIM}; space constraints preclude a
detailed discussion here. Fig.\ \ref{fig:CDF_schematic} shows schematically
one-quarter of the spectrometer, which surrounded the point (actually a region
about 0.5 m long) inside the Tevatron beam pipe at which the proton and
antiproton beams collided. 
Fig.\ \ref{fig:CDF_photo} shows
the actual layout. 
Fig.\ \ref{fig:Zmumu} is an {\em event display}, i.e., a schematic diagram showing 
the particle tracks as reconstructed by the spectrometer; the event shown 
contains a high-momentum muon ($\mu^-$) and antimuon ($\mu^+$) resulting from 
the production and decay of a $Z^0$ gauge boson. In the figure the beam axis 
runs into and out of the page, as does the magnetic field produced by the 
superconducting solenoidal
momentum-analyzing electromagnet. The curvature of charged-particle tracks due 
to the magnetic field is clearly evident.  A positive 
identification of a muon can be made for those trajectories that
pass through the massive hadronic calorimeter relatively unscathed and
leave signals in surrounding scintillators and wire chambers.

\begin{figure}[htb]
\hspace{-.125in}
\centerline{
\includegraphics[width=.5\linewidth,trim=20 0 0 0mm,clip]{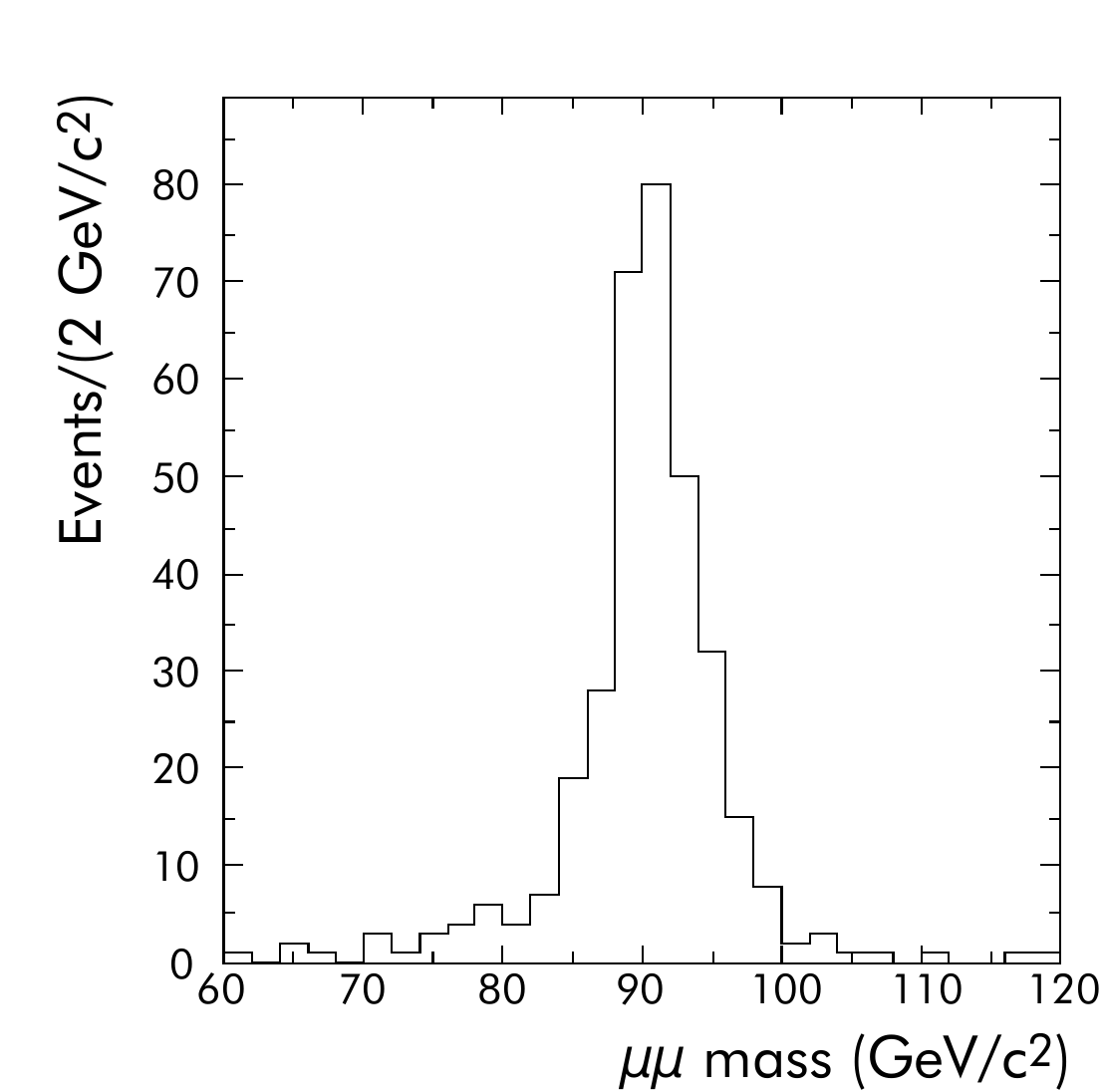}}
\caption[Dimuon mass spectrum obtained by the CDF collaboration]{Dimuon mass spectrum obtained by the CDF collaboration.\label{fig:Z}}
\end{figure}

Fig.\ \ref{fig:Z} shows the distribution in muon-pair 
mass observed by CDF. The prominent peak 
at 91\,GeV/$c^2$ is due to the 
$Z^0$. 
Its width
reflects both the (classical) measurement resolution of the magnetic
spectrometer and the (quantum-mechanical) uncertainty on the
$Z^0$'s mass (intrinsic width $\Gamma=2.49$\,GeV/$c^2$ fwhm) due to its
short lifetime. Since $\Gamma c^2$ represents the $Z^0$'s energy uncertainty, 
and the lifetime $\tau$ its duration uncertainty, they satisfy
a version of the Heisenberg uncertainty relation: $\Gamma c^2\tau=\hbar/2$.

\subsection{Double Chooz Experiment}
The Double Chooz Experiment~\cite{Abe:2011fz}
provides a convenient example of a current reactor-antineutrino detector design, with two identical detectors 
situated in underground laboratories
1\,km and 340\,m from  the ``Chooz B'' power reactors in northern France. The goal of the experiment is to measure neutrino oscillation parameters via  very
precise differential measurement of the reactor antineutrino flux and energy spectrum.
Detector design and siting are driven by the need to reduce
backgrounds from cosmic rays and radioactivity, which otherwise would
overwhelm the desired neutrino signal.  For example, one of the Double Chooz
detectors typically observes a neutrino rate of $\sim$\,5\,mHz, with a background
rate of 120\,Hz prior to event filtering.  The 
underground siting reduces the background from
cosmic rays.

The antineutrinos are detected via the previously mentioned 
inverse $\beta$-decay interaction on protons,
\begin{equation}
        \bar{\nu}_e + p \rightarrow n + e^{+},
\end{equation}
where the positron kinetic plus annihilation energy spans a range from 1.022 to 7\,MeV, peaking at $\approx$\,3\,MeV, and the neutron kinetic energy is
typically $\sim$\,20\,keV. Each Double Chooz detector is designed to have
a single homogeneous target medium in order to minimize efficiency variations that
would reduce  measurement precision. 

\begin{figure}[htb]
\centerline{\includegraphics[width=6.5in,trim=20 0 20 0mm,clip]{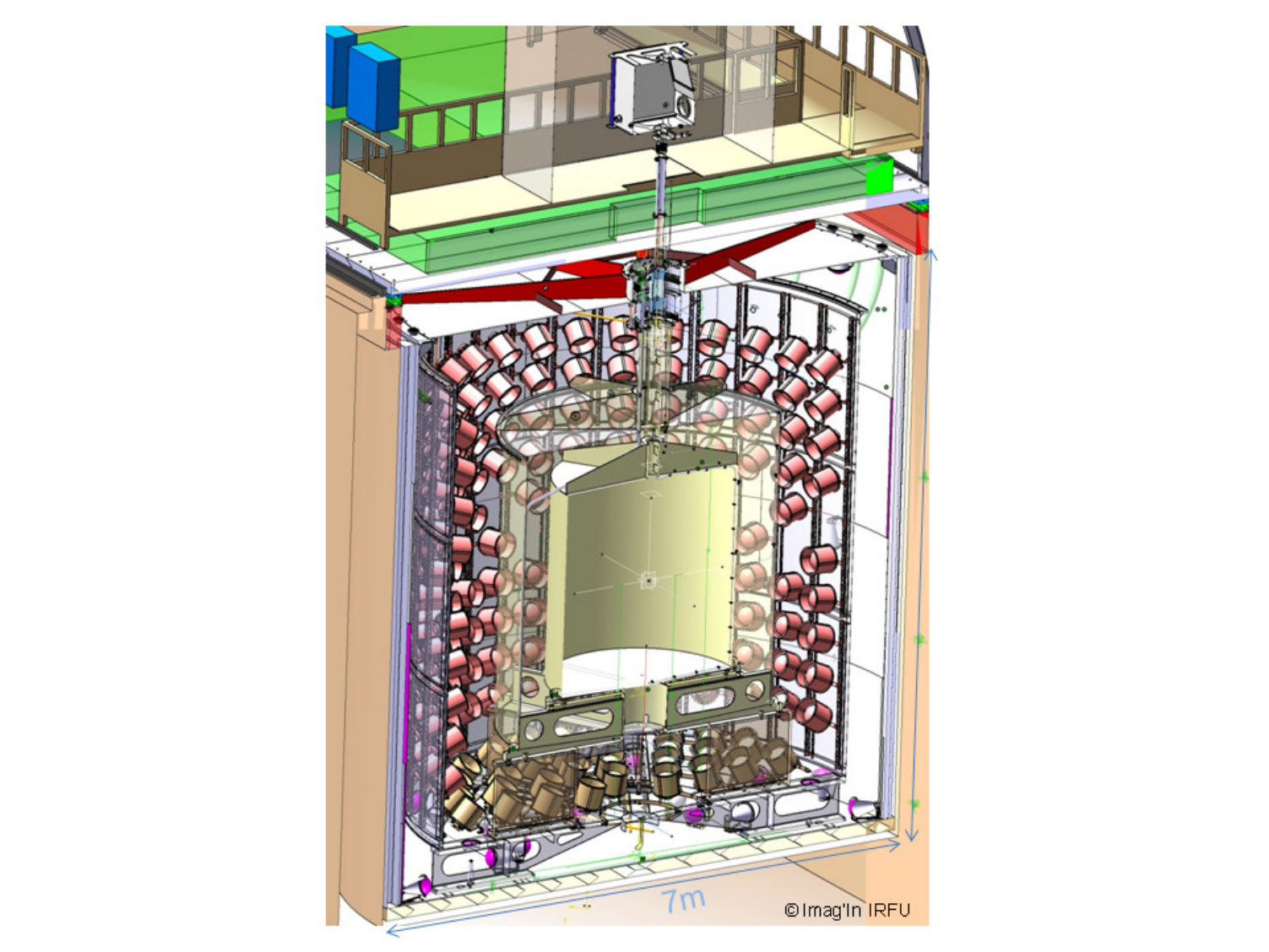}}
\caption[Cutaway view of a Double Chooz detector]{
Cutaway view of a Double Chooz reactor-neutrino detector. The inner
volume contains an organic liquid scintillator loaded with 0.1\%
gadolinium for neutron capture, surrounded by an unloaded ``gamma catcher''
scintillator, itself surrounded by a non-scintillating buffer liquid to shield
against radioactive backgrounds. The scintillators are contained in transparent
acrylic vessels and viewed by photomultiplier tubes. Cosmic-ray-induced backgrounds 
are suppressed using a surrounding liquid-scintillator ``inner veto" volume and 
an overhead ``outer veto" employing multiple layers of scintillation hodoscopes read out with wavelength-shifting fibers and multi-anode phototubes.
}
\label{fig:DC_cutaway}
\end{figure}

Figure~\ref{fig:DC_cutaway}
schematically illustrates the  Double Chooz detector design.  At the center of the
detector is a 10\,m${}^3$ volume of organic liquid scintillator containing  0.1\%
of a dissolved  organic gadolinium compound.  A large (atomic) fraction of the
scintillator is hydrogen, providing target protons, while  gadolinium has
a very large neutron capture probability,  such events being signaled by emission of gamma rays totaling
$\approx$\,8\,MeV of energy. Neutrons not captured on gadolinium are captured by hydrogen nuclei, with emission of $\approx$\,2\,MeV of gamma-ray energy. Neutrino interactions in the target volume result in
a positron that slows, stops, and annihilates with an electron to produce
two 0.511\,MeV gamma rays.  The positron kinetic energy and annihilation
gamma rays (when converted, by Compton scattering or photoelectric processes) cause
prompt scintillation-light emission in an amount proportional to
the total energy deposited in the scintillator.   The neutron from a 
neutrino interaction scatters repeatedly from protons, slowing to thermal
energies in $\approx$\,0.5\,$\mu$s, then diffuses randomly until captured by
gadolinium or hydrogen, with a capture lifetime of $\approx$\,25\,$\mu$s.
The gamma rays from the neutron capture then also are converted, resulting
in further scintillation light. 

The detector as a whole thus acts as a calorimeter,  the
signature of a neutrino event being a prompt light  pulse from the 
target scintillator,
with a variable amplitude proportional to the original neutrino energy
minus 0.8\,MeV, and a delayed light pulse some tens of microseconds
later, corresponding to 8 or 2\,MeV energy. The criteria of two pulses, each
within appropriate energy ranges, and with a time delay between them
appropriate for a neutron capture, helps to suppress background signals.

The target scintillator is contained in a transparent acrylic vessel, to allow 
scintillation light to escape, and surrounded by a (so-called) ``gamma catcher'' volume 
of liquid scintillator without gadolinium; this allows the efficient 
detection of gamma rays that escape from the inner target volume.
The gamma catcher is itself enclosed in a transparent acrylic vessel,
and surrounded by non-scintillating organic ``buffer'' liquid, which 
helps shield the inner volumes from gamma-ray backgrounds from the
outside of the detector.

The scintillation light from the target and gamma catcher (``inner detector'') scintillators
is detected by 380 ten-inch-diameter hemispherical PMTs  mounted on the inside of the steel tank that encloses the inner 
detector volumes.  Outside of that tank, a further 90 PMTs are immersed in liquid scintillator,  used to veto any cosmic-ray
backgrounds that enter the detector, with a muon tracker system composed of  plastic scintillation hodoscopes located
on top of the detector to assist in characterizing and vetoing 
cosmic-ray backgrounds.  The detectors were built into pits that were
excavated from the laboratory floors, and surrounded  with  iron shielding. 

\subsubsection{Backgrounds}
The 8\,MeV energy from neutron capture on gadolinium is 
well above those of natural radioactive backgrounds, so the primary source of
false signals (backgrounds) in the Double Chooz detectors is 
high-energy cosmic-ray interactions in the detectors or nearby material 
that either produce neutrons (by cosmic-ray interaction with nuclei) or
yield unstable nuclei in the detector that decay with sufficient energy
to mimic a neutron capture. 
One such nucleus is ${}^8$Li, produced by nuclear spallation from
muon interactions.  When a cosmic-ray muon passes through the detector,
it deposits a large amount of energy and produces a very large 
scintillation signal; however the signal is so large and variable that
there is no indication whether a spallation occurred.
${}^8$Li has a long enough  half-life (840\,ms) that one
cannot correlate its decay to the passage of  a particular cosmic ray, and with a beta-decay
energy of 13\,MeV,  some fraction of its decays will be within
the energy range of neutron captures.

One of the
more powerful techniques used by Double Chooz and other reactor-neutrino experiments is to compare the detected rate of candidate
neutrino events to the reactor power output: the neutrino rate is
proportional to the power, while cosmic-ray and background radioactivity
give a constant rate of background events.  However, reactor-neutrino
experiments at short distances, such as inside reactor containment buildings,  tend to suffer from
``reactor-related'' neutron and gamma backgrounds that are proportional
to  reactor power.  

\subsubsection{Neutrino Results}

\begin{figure}
\centerline{\includegraphics[width=.75\columnwidth]{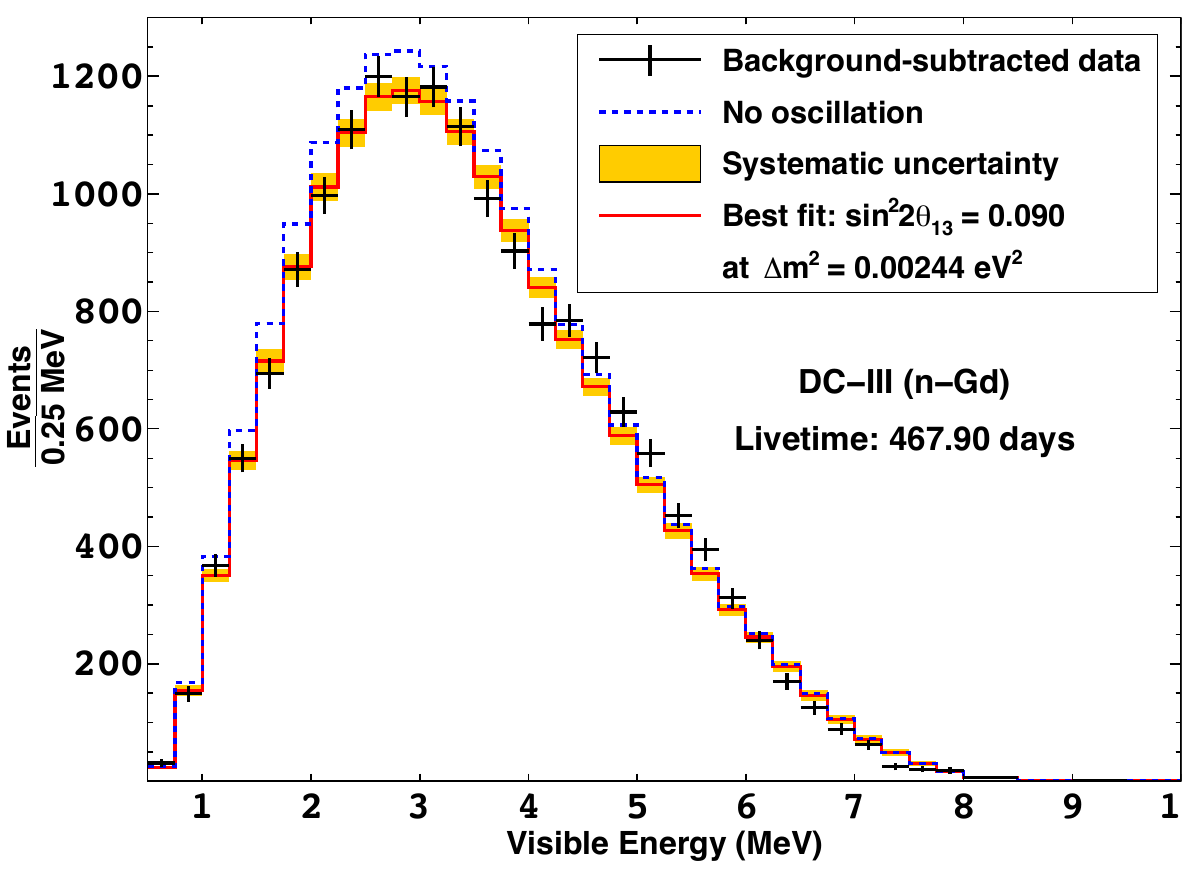}}
\caption[Prompt energy spectrum of neutrino events in the Double
Chooz experiment.]{Prompt energy spectrum of neutrino events in the Double
Chooz experiment. The spectrum is distorted by oscillation of the
antineutrinos to other neutrino flavors over the 1\,km distance
between their production in a reactor core and the Double Chooz
detector.}\label{fig:DCprompt} 
\end{figure}

Figure~\ref{fig:DCprompt} shows the ``prompt'' energy spectrum measured for neutrinos at Double Chooz~\cite{Abe:2014bwa}, with backgrounds from fast neutrons and spallation products subtracted. The suppression of the spectrum between 1.5 and 4\,MeV is caused by oscillation of electron antineutrinos to other flavors of neutrino.

\section{{Summary}}

Following an introduction to particle physics and particle detectors, we have
considered three contrasting examples of subatomic-particle spectrometers, ranging
from the relatively simple (Double Chooz and HyperCP) to the complex (CDF). While the brief
discussion just given illustrates the variety of issues encountered in
designing particle spectrometers and their electronic instrumentation, the
actual design process is quite involved. Extensive computer simulation is
generally employed to tailor a design optimized for the problem at hand.
Requirements for performance and reliability often come up against practical
constraints on cost and on development and assembly time. 

The ongoing development of new technology for particle detectors and their 
instrumentation, together with the
development of increasingly intense and energetic particle beams, 
make measurements possible that were previously not feasible.
When new detector technology is employed, simulation studies 
must be combined with prototype tests both on the bench and at test beams.
The investigation of matter and energy at ever deeper and more 
sophisticated levels exemplifies fruitful collaboration among
scientists and engineers.

\section*{{Acknowledgements}}
\addcontentsline{toc}{section}{Acknowledgements}

The authors thank the Double Chooz and Fermilab CDF and HyperCP collaborations for
permission to reproduce their results.

\section*{{Reading List}}

Good introductory treatments of special relativity and particle physics may be
found in textbooks on modern physics, for example, 
\begin{quote}
A. Beiser, {\it Concepts of
Modern Physics}, 6th ed., New York: McGraw-Hill, 2002.

K. S. Krane, {\it Modern
Physics}, 3rd ed., New York: John Wiley \& Sons, 2012.

H. C. Ohanian, {\it Modern
Physics}, 2nd ed., Engelwood Cliffs, NJ: Prentice Hall, 1995.
\end{quote}
There are also more elementary and abbreviated treatments in general-physics 
textbooks, for example, 
\begin{quote}
D. Halliday, R. Resnick, and J. Walker, {\it Fundamentals 
of Physics}, 10th ed., New York: John Wiley \& Sons, 2013.

H. C. Ohanian, {\it Physics}, 2nd ed., Vol.\ 2, Expanded, New York: Norton,
1989.
\end{quote}
Introductory texts on particle physics include 
\begin{quote}
D. J. Griffiths, {\it Introduction to Elementary Particles}, 2nd ed., New York: Harper
and Row, 2008.

D. H. Perkins, {\it Introduction to High Energy
Physics}, 4th ed., Cambridge, UK: Cambridge Univ.\ Press, 2000.
\end{quote}

Detailed treatments of particle detection techniques may be found in 
\begin{quote}
K. A. Olive et al., {\it Review of Particle Physics} ({\it op cit.} 
\cite{PDG}).

T. Ferbel (ed.), {\it
Experimental Techniques in High Energy Nuclear and Particle Physics}, 
2nd ed., Singapore: World Scientific, 1991.

W. R. Leo, {\it Techniques for
Nuclear and Particle Physics Experiments}, 2nd ed.,
New York: Springer, 1994.
\end{quote}

\end{document}